\documentclass[preprint,12pt]{elsarticle}
\usepackage{graphicx}
\usepackage{amsmath}
\usepackage{amsfonts}
\usepackage{multirow}
\usepackage{physics}
\usepackage{bm}
\usepackage[colorlinks, linkcolor=blue, anchorcolor=blue, citecolor=blue, urlcolor=blue]{hyperref}

\journal{Physica B: Condensed Matter}

\begin{document}

\title{Shear Viscosity at the van Hove singularity}
\author[inst1]{Yi-Hui Xing\corref{cor1}}
\ead{xingyihui20@iphy.ac.cn}
\author[inst1]{Wu-Ming Liu\corref{cor1}}
\ead{wliu@iphy.ac.cn}
\address[inst1]{Beijing National Laboratory for Condensed Matter Physics, Institute of Physics, Chinese Academy of Sciences, Beijing, 100190, China}
\cortext[cor1]{Corresponding author}

\begin{abstract}
Boltzmann transport theory may fail in strongly correlated systems when quasiparticles are poorly defined. Using a diagrammatic Kubo approach within a large-$N$, large-$N'$ random-flavor model, we study transport near a van Hove singularity. While the conductivity agrees with the Boltzmann prediction, the dc shear viscosity differs qualitatively in the dirty quantum-critical regime, where impurity broadening invalidates the sharp-fermion approximation. In the clean regime, the optical shear viscosity scales as
$\mathrm{Re}\,\eta(\Omega,T)\sim (|\Omega|^{3/2}+T^{3/2})/\Omega^2.$
In the dirty regime, the full dc viscosity is
$\eta_{\mathrm{dc}}(T)=\eta_0-AT,$
where \(\eta_0\) is the temperature-independent impurity background and
\(A\) is a nonuniversal coefficient that depends on the impurity
scattering rate and the ultraviolet cutoffs of the VHS patch theory.
These results establish shear viscosity as a sensitive probe of non-Boltzmann transport and critical scattering mechanisms.
\end{abstract}


\maketitle

\section{Introduction.}
\label{1}
For the study of transport properties in metals, the diagrammatic Kubo approach and the linearized Boltzmann-equation constitute two of the most commonly used and mutually complementary theoretical frameworks \cite{RevModPhys.58.323, Maslov_2017}.
In the quasiparticle regime, it is well established that the Kubo formulation reduces to a linear integral equation equivalent to the linearized Boltzmann-equation, and that the two approaches yield identical leading-order transport coefficients within this controlled regime \cite{HOLSTEIN1964410,PhysRevB.27.5775,PhysRevLett.97.096604}.
However, in the quantum critical regime where the Landau quasiparticle picture breaks down, non-Fermi-liquid behavior \cite{RevModPhys.79.1015} emerges together with anomalous transport properties, such as those observed in strange metals \cite{science.abq6011,Greene2020,Cao2020}.
A natural question then arises: do the diagrammatic Kubo approach and the linearized Boltzmann-equation remain equivalent in this regime?
In this article, we show that the answer remains affirmative as long as the fermion spectral function remains sharply peaked at the Fermi surface.
However, when the fermions acquire a large broadening compared with the boson modes, the two approaches generally lead to different conclusions.
We illustrate this distinction explicitly using the shear viscosity \cite{landau2013course,PhysRevE.108.044101} as a representative example.

Shear viscosity characterizes the transverse diffusion of momentum density and constitutes a central transport quantity in the hydrodynamics.
When electron-electron collisions dominate, the electronic shear viscosity controls momentum dissipation in the electronic fluid and thus plays an important role in electrical transport experiments \cite{2fcg-zmwt,science.aaf2487,PhysRevLett.113.235901}.
Experimentally, materials such as graphene \cite{science.aad0201,science.aad0343,levitov2016electron,PhysRevB.95.075127} and ruthenates \cite{PhysRevLett.86.5986,PhysRevLett.106.096401,hunter2025} have shown compelling evidence for electronic fluid behavior, including viscous drag arising from hydrodynamic electron flow. 
This behavior is often discussed in regimes where the quasiparticle picture is not the most appropriate description of transport.
In a broader class of solid-state systems, Corbino disk geometries \cite{PhysRevLett.113.235901} and sound-attenuation measurements \cite{2fcg-zmwt} are widely employed to extract symmetry-resolved shear viscosities.

We focus on the shear viscosity of a Fermi surface hosting a van Hove singularity (VHS) and coupled to critical Ising-type boson fluctuations, which destroy the quasiparticle nature of the fermions. 
In our previous work \cite{xing2025strangemetal}, we showed that this mechanism can account for strange metal behavior. 
Experimentally, it has also been observed that when the system is tuned away from the VHS, the strange metal behavior disappears and the system crosses over to a conventional Fermi liquid \cite{pnas.2321665121,Greene2020,Ghiotto2021}. 
As summarized in Table~\ref{tab:scaling}, the shear viscosity exhibits qualitatively different scaling behaviors in the clean, dirty, and random-Yukawa regimes.
In the clean limit, we find the optical shear viscosity scaling $(|\Omega|^{3/2}+T^{3/2})/\Omega^2$. 
Within our large-\(N\) analysis, strong impurity scattering broadens the fermion spectral function, eliminating its sharp peak at the Fermi surface. 
Consequently, after subtracting the dominant temperature-independent background arising from impurity scattering, the leading temperature-dependent contribution to the dc shear viscosity deviates from the Boltzmann prediction \(T^{3/2}/\Gamma^2\) and instead scales as
$
T\left(\Lambda_U^2+e^{2\Lambda}\Gamma\right)/\Gamma^2,
$
where \(\Lambda_U\) and \(\Lambda\) denote the ultraviolet cutoffs defined below.
By contrast, the temperature-dependent contribution to the dc conductivity remains linear in \(T\) in both the Boltzmann and diagrammatic Kubo treatments.

Upon introducing spatially random fluctuations into an otherwise uniform boson-fermion coupling, strange metal behavior can also emerge away from the VHS \cite{science.abq6011,PhysRevB.105.235111}. 
We find that such a spatially random Yukawa interaction gives rise to a shear viscosity scaling as $(|\Omega|+T)/\Omega^2$, in contrast to the conductivity, for which the two mechanisms yield the same result. 
This qualitative difference suggests that shear viscosity may serve as an experimental diagnostic for determining whether critical boson–fermion coupling at the VHS provides an appropriate description of some strange metals.


\begin{table}
\caption{Summary of the leading shear-viscosity scalings obtained in this work in the perturbative large-$N$ and large-$N'$ random-flavor frameworks.
Note that we present only the leading frequency and temperature dependence, 
their coefficients are generally nonuniversal and
may depend on the ultraviolet cutoffs of the VHS patch theory.
In particular, in the dirty limit, the dominant impurity-scattering contribution to the dc shear viscosity is a temperature-independent constant of order $N\Lambda_U^4/\Gamma$, which is not shown.}
\label{tab:scaling}
\centering
\begin{tabular}{c c c c}
Regime &  Leading scaling & Relevant equation \\
\hline
$|\Omega|\gg T\gg \Gamma$ &  $|\Omega|^{-1/2}$ & Eq.~\eqref{ov-T01} \\
$T\gg |\Omega|\gg \Gamma$ &  $T^{3/2}/\Omega^2$ & Eq.~\eqref{alaa} \\
$\Gamma \gg T,\ \Omega=0$ &  $ T^{3/2}/\Gamma^2$ & Eq.~\eqref{dcsv1} \\
$\Gamma \gg T,\ \Omega=0$ &  $T(\Lambda_U^2+e^{2\Lambda}\Gamma)/\Gamma^2$ & Eq.~\eqref{dcsv2} \\
Spatially random Yukawa, $\Gamma=0$ & $(|\Omega|+T)/\Omega^2$ & Sec.~\ref{6} \\
\end{tabular}
\end{table}

The rest of this paper is organized as follows.
In Sec.~\ref{2}, we introduce the low-energy model and the critical boson propagator near the VHS.
In Sec.~\ref{3} and ~\ref{4}, we define the shear-stress form factor used in the patch theory and derive the optical shear viscosity in the clean limit.
In Sec.~\ref{5}, we analyze the dirty dc limit and explain why the Prange--Kadanoff or Boltzmann reduction fails once impurity broadening dominates the fermion spectral function.
In Sec.~\ref{6}, we discuss spatially random Yukawa interactions and summarize the physical implications.
Several technical points are deferred to the Appendices: Appendix~\ref{A} derives the diagrammatic expression for the current correlator; Appendix~\ref{B} summarizes the momentum integrals used in different limits; Appendix~\ref{C} discusses the relation to the Boltzmann equation; Appendix~\ref{D} shows that impurity ladder corrections to the Yukawa vertex are subleading in the large-$N$ random-matrix disorder ensemble used in the main text; Appendix~\ref{E} explains that if one nevertheless dresses the fermion propagator, the corresponding vertex correction must be treated consistently, but this does not modify the leading scaling results obtained in the perturbative large-$N/N'$ expansion.

\section{The Model.}
\label{2}
We focus on the low-energy effective theory describing fermions near the VHS saddle point coupled to critical Ising-nematic boson fluctuations, which reads
\begin{equation}
\begin{aligned}
	\mathcal{L}=&\sum_{m}\psi^\dagger_{m}(\partial_\tau-\partial_x^2+\partial_y^2)\psi_{m}+\frac{1}{2}\sum_l\phi_l(\Delta -\partial_\tau^2 - \bm{\nabla}^2 )\phi_l +\\
	&\frac{1}{N}\sum_{mn}w_{mn}({\bm x})\psi^\dagger_m\psi_n+\frac{1}{\sqrt{NN'}}\sum_{mnl}J_{mnl}\psi^\dagger_{m}\psi_{n}\phi_l+h.c.,		
\end{aligned}
		\label{eff}
\end{equation}
where $m,n=1,\ldots,N$ and $l = 1,\ldots,N'$ label the flavor indices of the fermion and boson fields, respectively. The interaction $w_{mn}({\bm x})$ represents impurity scattering and is assumed to obey a Gaussian distribution in both real space and flavor space, whereas $J_{mnl}$ denotes a spatially uniform Yukawa interaction that is Gaussian distributed only in flavor space\cite{xing2025strangemetal,science.abq6011}.
The parameter $\Delta$ is the bare boson mass.
At the quantum critical point, the renormalized boson mass vanishes; equivalently, $\Delta-\Pi_{\rm J}(0,0)=0$ in our convention below, where $\Pi_{\rm J}$ denotes the boson self-energy.
Thus $\Delta$ is retained in the microscopic action but drops out of the critical damping part of the boson propagator.

The Ising-nematic field considered here describes a $B_{1g}$ Pomeranchuk instability in a tetragonal lattice. At the lattice level, the corresponding Yukawa coupling carries the form factor
\[
 f_{B_{1g}}({\bm k})=\cos k_x-\cos k_y.
\]
For the cuprate and ruthenate band structures motivating our discussion, the relevant VHSs occur at the Brillouin-zone boundary, near $X=(\pm\pi,0)$ and $Y=(0,\pm\pi)$. The form factor is approximately constant within each VHS patch,
\[
 f_{B_{1g}}(X)=-2+O(q^2),\qquad
 f_{B_{1g}}(Y)=+2+O(q^2).
\]
The momentum-independent Yukawa vertex in Eq.~\eqref{eff} should therefore be understood as the local projection of this $B_{1g}$ form factor within a given VHS patch, with the opposite signs restored between the symmetry-related $X$ and $Y$ patches.

In our previous work \cite{xing2025strangemetal}, we showed that effective theory \eqref{eff} accounts for the linear-in-temperature resistivity and the $T\ln T$ electronic specific heat of the strange metal in the quantum critical regime. This behavior is also consistent with experimental observations in several material systems, such as cuprate superconductors \cite{Greene2020}, twisted bilayer systems \cite{Cao2020,Jaoui2022,Ghiotto2021}, and ruthenates \cite{pnas.1915224117}, where strange metal phases emerge in a finite-temperature fan emanating from doping near a VHS.
All of these striking behaviors can be traced back to the singular damping term of the critical boson fluctuations:
\begin{equation}
	\Pi_{\rm J}(i\Omega_m,\bm{q})-\Pi_{\rm J}(0,0)=-\frac{|J|^2N|\Omega_m|}{4\pi N'|\epsilon_{\bm q}|}f(\frac{\Gamma}{|\epsilon_{\bm q}|}),
	\label{bose}
\end{equation}
where $\epsilon_{\bm q}=q_x^2-q_y^2$, $\Gamma=|w|^2\ln\Lambda_U/2\pi$ denotes the impurity scattering rate, $\Lambda_U$ is the ultraviolet momentum cutoff of the VHS patch theory and $f(\Gamma/|\epsilon_{\bm q}|)$ is an impurity controlled function (see Appendix \ref{B}). In the clean limit $\Gamma\to 0$, $f(x)\to 1$, and Eq.\eqref{bose} reduces to the standard Landau damping form. In contrast, in the dirty limit $\Gamma \to \infty$, $f(x)\to \ln(4ex)/2\pi x$, so that the boson self-energy given by Eq.\eqref{bose} becomes inversely proportional to $\Gamma$. A similar behavior also occurs for a generic convex Fermi surface \cite{science.abq6011}.
As discussed in Appendix \ref{D}, the impurity ladder associated with the same disorder is subleading in the large-$N$ expansion for the random off-diagonal Yukawa bilinear considered here.
Therefore Eq.~\eqref{bose} is the leading large-$N$ dirty polarization kernel, rather than the diffusive response of a conserved flavor-singlet density.

Here, the terms “clean” and “dirty” denote asymptotic scale hierarchies
relative to the impurity scattering rate, rather than the complete absence
or presence of interactions. In the clean optical regime, the impurity rate
is the smallest infrared scale,
$
\Gamma\ll\min(T,|\Omega|),
$
with the \(T=0\) condition understood as \(\Gamma\ll|\Omega|\). We may
therefore set \(\Gamma=0\) before evaluating the finite-frequency response.
The critical boson-fermion interaction remains present and generates the
regular optical viscosity.
In the dirty regime,
$
\max(T,|\Omega|,\gamma_{\mathrm{int}}(T))\ll\Gamma,
$
where
$
\gamma_{\mathrm{int}}(T)
\equiv-\operatorname{Im}\Sigma_{\mathrm{int}}^R(\omega\sim T)
$
denotes the interaction-induced fermionic width. In this hierarchy,\(\Gamma\) must be retained in the fermion propagators, but the external
frequency may be taken to zero at fixed \(\Gamma\), giving the
impurity-dominated dc response. Consequently, the limits
\(\Gamma\to0\) and \(\Omega\to0\) are not interchangeable.

We turn to a different transport quantity, the shear viscosity, motivated by two considerations. 
First, we will show that, at the VHS, the conductivity exhibits the same temperature dependence in both the clean and the dirty limits; this agreement is in fact accidental. 
Using the shear viscosity as an example, we demonstrate that such a coincidence need not hold in general. 
Second, Ref.~\cite{science.abq6011} pointed out that spatially random interactions can also produce strange metal, including a linear-in-$T$ resistivity and an optical conductivity scaling as $1/|\Omega|$. 
In contrast, for the optical shear viscosity, spatially uniform and spatially random interactions lead to qualitatively different behaviors. 
This sensitivity provides a diagnostic to distinguish which mechanism plays the dominant role in driving the strange metal.

\begin{figure}
		\centering
		\includegraphics[width=\columnwidth]{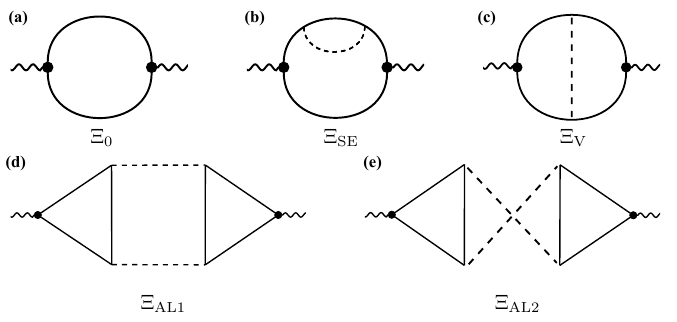}
		\caption{Leading-order Feynman diagrams for the momentum density current correlation function. (a) The one-loop diagram $\Xi_{0}$; (b) the self-energy diagram $\Xi_{\rm SE}$; (c) the vertex diagram $\Xi_{\rm V}$; and (d,e) the two Aslamazov–Larkin diagrams $\Xi_{\rm AL1}$ and $\Xi_{\rm AL2}$. Solid lines, dashed lines, and filled black circles denote fermion propagators, boson propagators, and current vertices, respectively.
		}
		\label{Fig1}
	\end{figure}

We compute the shear viscosity within perturbation theory using the Kubo formula, ${\rm Re}[\eta(\Omega)]=-{\rm Im}[\Xi(i\Omega_m\to \Omega+i0^+)]/\Omega$, which reduces the problem to evaluating the set of Feynman diagrams shown in Fig.~\ref{Fig1}.
Here, $\Xi(i\Omega_m)$ represents the correlation function of the uniform momentum  density current $T_{xy}=\sum_{{\bm k},m}t_{\bm k}\psi^\dagger_{\bm k,m} \psi_{\bm k,m}$, with $t_{\bm k}$ identified as the deformation potential.
In a microscopic lattice model, the shear-stress vertex is not universal and should be obtained by differentiating the lattice Hamiltonian with respect to shear strain,
$\gamma_{xy}({\bm k})=\partial \epsilon({\bm k};\epsilon_{xy})/\partial \epsilon_{xy}|_{\epsilon_{xy}=0}$.

The symmetry distinction between the critical mode and the shear probe is important. The critical field $\phi$ transforms as $B_{1g}$, whereas the shear strain $\epsilon_{xy}$ transforms as $B_{2g}$. Since $B_{1g}\otimes B_{2g}=A_{2g}$, a uniform linear coupling $\epsilon_{xy}\phi$ is forbidden in the tetragonal symmetric phase. The lowest local, uniform, nonderivative mixed invariant is
$
 \delta\mathcal{L}_{\phi\text{-}\epsilon}
 =\frac{\lambda}{2}\epsilon_{xy}^{2}\phi^{2}.
$
This term only shifts the boson mass, $\Delta_{\rm eff}=\Delta+\lambda\epsilon_{xy}^{2}$. It does not enter the linear $B_{2g}$ response about $\epsilon_{xy}=0$; at finite background strain, criticality is defined by tuning the renormalized mass to zero. This statement concerns the $B_{2g}$ channel considered here; a $B_{1g}$ strain can, in general, couple linearly to a $B_{1g}$ order parameter.

The leading linear coupling of the $B_{2g}$ shear strain is therefore to the electronic band. This can be seen explicitly from the diagonal hoppings of a square lattice. Writing their contribution to the dispersion as
$
 \epsilon_{\rm diag}({\bm k};\epsilon_{xy})
 =-2t_{+}\cos(k_x+k_y)-2t_{-}\cos(k_x-k_y),
 \
 t_{\pm}=t'\pm g_{xy}\epsilon_{xy},
$
the electronic shear vertex becomes
$
 \gamma_{xy}({\bm k})
 =\left.\frac{\partial\epsilon_{\rm diag}({\bm k};\epsilon_{xy})}
 {\partial\epsilon_{xy}}\right|_{\epsilon_{xy}=0}
 =4g_{xy}\sin k_x\sin k_y.
$
Near $X=(\pi,0)$, $\sin(\pi+q_x)\sin q_y=-q_xq_y+O(q^4)$, and the same low-energy structure is obtained near the symmetry-related $Y$ patch. After absorbing the nonuniversal coefficient and sign into the definition of the probe vertex and relabeling the patch momentum, this gives precisely $t_{\bm k}=k_xk_y$ used in our calculation. The viscosity computed below is consequently the electronic $B_{2g}$ shear response dressed by $B_{1g}$ critical fluctuations, rather than a response generated by a direct linear coupling between the $B_{2g}$ strain and the $B_{1g}$ boson.

We first provide a simple estimate based on scaling arguments. 
If the fermion spectral weight is assumed to sharply distributed near the Fermi surface, the transport coefficients are governed solely by the boson dynamical critical exponent $z$, and the boson self-energy \eqref{bose} renormalizes the dynamics to $z=4$.
The transport scattering rate relevant for the conductivity is determined by the boson density of states (${\rm DOS}\sim T^{d/z}$) together with the small angle scattering factor $1-\cos\theta\sim q^2$, which leads to a linear-in-temperature behavior, ${\rm Re}[\sigma]\sim {\rm DOS}\cdot q^2 \sim T$. 
The key difference for the shear viscosity lies in the small-angle scattering factor, which scales as $\sim q^{4}$. As a result, the shear viscosity scales as $T^{3/2}$, in agreement with the analytical result in Eq.\eqref{alaa}.

\section{Zero-temperature Optical Shear Viscosity.}
\label{3}
In the following, we concentrate on the ``optical" shear viscosity in the clean limit, corresponding to case $\Gamma \ll |\Omega|$. For simplicity, we neglect impurity scattering ($\Gamma=0$). We first analyze the low-temperature limit $|\Omega|\gg T$, and subsequently turn to the finite-temperature case $|\Omega|\ll T$.

We focus on the regime near the quantum critical point, where the boson mode acquires a self-energy and becomes damped. 
For the fermions, we still use the bare propagator, $G(i\omega,{\bm k})=(i\omega_n-\epsilon_{\bm k})^{-1}$. 
This is the leading nontrivial contribution in the combined perturbative expansion in \(|J|\) and the large-\((N/N')\) limit.
In this counting, the fermion self-energy and the associated current-vertex corrections are parametrically subleading relative to the boson self-energy retained in Eq.~\eqref{bose}, and we therefore use bare fermions in the leading diagrams.
A fully dressed fermion propagator would require the corresponding vertex correction to be included consistently; this point is discussed in Appendix~\ref{E}.

At the one loop level, the momentum density current correlator is given by $\Xi_0(i\Omega_m)\approx -N\int \frac{d^2{\bm k}}{(2\pi)^2}t_{\bm k}^2\delta(\epsilon_{\bm k})$.
The result is a frequency independent constant, which gives no contribution to shear viscosity after analytic continuation.

Next, we consider the contributions from the higher loop Feynman diagrams shown in Fig.~\ref{Fig1}(b-e), whose sum is denoted by
\begin{equation}
	\begin{aligned}
	&\Xi(i\Omega_m)\\
	=&\frac{N^2|J|^4}{4N'\Omega_m^2}T\sum_{\Omega}\int \frac{d^2\bm{p}}{(2\pi)^2}\int_{\Omega,{\bm p}} D[{\bm k}_1]\int_{\Omega,{\bm p}} D[{\bm k}_2]\\
		 &\times (\Delta t)^2 D(i\Omega-i\Omega_m/2,\bm{p})D(i\Omega+i\Omega_m/2,\bm{p}).
	\end{aligned}
	\label{all}
\end{equation}
The explicit derivation is deferred to the Appendix \ref{A}. Here,
\begin{equation}
	\Delta t=t_{\bm{k}_2}-t_{\bm{k}_2+{\bm p}}+t_{{\bm k}_1+{\bm{p}}}-t_{{\bm k}_1}
	\label{delt}
\end{equation}
represents the change in the deformation potential before and after scattering, while
$\int_{\Omega,{\bm p}} D[{\bm k}]$ 
is a simplified notation defined in Appendix \ref{A}.

For the saddle-point dispersion at the VHS, the deformation potential satisfies $t_{\bm k}=k_xk_y$, so that $\Delta t = k_{1x}p_y+p_x k_{1y}-(k_{2x}p_y+p_x k_{2y})$ is nonzero. As a result, a nontrivial shear viscosity already arises at leading order. 
In contrast, for a generic convex Fermi surface dispersion $\epsilon_{\bm k}=v_{\rm F}k_\perp+k_\parallel^2$, the deformation potential takes $t_{\bm k}=v_{\rm F}k_\parallel$, which leads to $\Delta t = 0$. In this case, obtaining a nontrivial shear viscosity requires considering subleading contributions, such as effects due to deviations away from the Fermi surface. These corrections give rise to higher order power-law dependences on frequency and temperature.
These features are analogous to those encountered in conductivity calculations \cite{PhysRevB.109.115156,xing2025strangemetal,science.abq6011}, in which the deformation potential is replaced by the group velocity.

After applying the Prange--Kadanoff approximation \cite{PhysRev.134.A566}, the Green functions is reduced to a delta function localized on the Fermi surface. 
We then use the momentum-integration formulas \eqref{delintk} in Appendix \ref{B} to evaluate the fermionic momentum integrals. 
This gives the total momentum-density-current correlation function at zero temperature as
\begin{equation}
	\begin{aligned}
		&\Xi(i\Omega_m)\\
		=&\,
		\frac{N|J|^2}{16\pi^2\Omega_m^2}
		\int_0^{|\Omega_m|}d\Omega\,
		\bigl(|\Omega_m|-\Omega\bigr)
		\int\frac{d^2p}{(2\pi)^2}
		\frac{p^4}{|\epsilon_{\bm p}|}
		D(i\Omega,\bm p)
		\\
		\sim&\,
		\frac{N|J|^2}{\Omega_m^2}
		\int_0^{|\Omega_m|}d\Omega\,
		\bigl(|\Omega_m|-\Omega\bigr)
		\int_{|\Omega|}^{\Lambda_U^2}dp
		\int_0^\Lambda d\theta\,
		\cosh^2(2\theta)
		\\
		&\times
		\frac{p^2}
		{p^2\cosh(2\theta)+N|J|^2|\Omega|/N'}
		\\
		\sim&\,
		N|J|^2e^{2\Lambda}\Lambda_U^2
		-\frac{N^{3/2}|J|^3}{N'^{1/2}}
		e^\Lambda|\Omega_m|^{1/2}
		-N|J|^2e^{2\Lambda}|\Omega_m|.
	\end{aligned}
	\label{ov-T0}
\end{equation}
We have also verified that the same conclusion can be reached directly at zero temperature without invoking the Prange-Kadanoff approximation. We refer the reader to our previous work, Ref.~\cite{xing2025strangemetal}, where several independent methods were shown to yield consistent results for the bosonic contribution involving $\int_{\Omega,\bm p}D(\Omega,\bm p)$.

Here and below, the momentum integrals are evaluated in hyperbolic coordinates. The hyperbolic-angle integral develops a strong divergence near the Fermi surface, $p_x=\pm p_y$, which is regulated by the cutoff $\Lambda$, whereas $\Lambda_U$ denotes the ultraviolet momentum cutoff. The lower bound $p\sim|\Omega|$ arises because, in the clean limit, the Landau-damping form of the boson propagator is valid only in the static regime, $|\Omega|\ll|\epsilon_{\bm p}|$ (see also Appendix \ref{B}). Although the first term in Eq.~\eqref{ov-T0} is the most divergent, it is independent of the external frequency and therefore does not contribute to the shear viscosity. Moreover, in the low-frequency regime
$
|\Omega_m|\ll N|J|^2e^{-2\Lambda}/N',
$
the third term is subleading to the nonanalytic $|\Omega_m|^{1/2}$ term. After analytic continuation and application of the Kubo formula, we therefore obtain the zero-temperature shear viscosity in the clean limit:
\begin{equation}
	{\rm Re}\,\eta(T\ll|\Omega|)
	\sim
	\frac{N^{3/2}|J|^3e^\Lambda}{N'^{1/2}}
	\frac{1}{|\Omega|^{1/2}}.
	\label{ov-T01}
\end{equation}
This contribution arises entirely from the higher-loop Feynman diagrams shown in Fig.~\ref{Fig1}; importantly, these diagrams do not cancel among themselves.

Finally, we comment on what would be obtained if the fermion propagator in Fig.~\ref{Fig1} were replaced by its fully renormalized form. 
At the quantum critical point, the fermion and boson sectors are mutually renormalized. 
For the VHS saddle-point dispersion, both the large-$N$ saddle-point analysis and the Eliashberg approach give
$
G^{-1}(i\omega,{\bm k})
=
i|J|\sqrt{\frac{N'}{N}}\,{\rm sgn}(\omega)|\omega|^{1/2}
-\epsilon_{\bm k}
$
\cite{xing2025strangemetal}.
If this propagator is inserted into the one-loop diagram in Fig.~\ref{Fig1}(a), one obtains
\begin{equation}
\begin{aligned}
	{\rm Re}[\eta(T\ll |\Omega|)]
	\sim
	\frac{N^{1/2}\Lambda_U^4}{N'^{1/2}|J|}
	\frac{1}{|\Omega|^{1/2}},
\end{aligned}
\label{ol-re}
\end{equation}
where $\Lambda_U$ is the momentum cutoff. 
The power-law dependence in Eq.~\eqref{ol-re} is the same as that in Eq.~\eqref{ov-T0}. 
However, this agreement should not be interpreted as evidence that the one-loop dressed bubble captures the transport physics. 
Rather, it is accidental: the one-loop diagram does not contain the small-angle scattering factor, and therefore the result is controlled by the single-particle decay rate, $1/\tau\sim |\Omega|^{1/2}$ \cite{PhysRevB.92.054305}. 

A similar situation occurs for the optical conductivity at the VHS. 
The dressed one-loop bubble gives a contribution scaling as
$\sim \Lambda_U^2/|\Omega|^{1/2}$, up to a different cutoff dependence. 
After the transport vertex structure, or equivalently the higher-loop diagrams in Fig.~\ref{Fig1}(b-e), is included, the frequency dependence is changed to $\sim 1/|\Omega|$. 
Thus the dressed one-loop result alone is not a reliable transport result.

For the higher-loop diagrams in Fig.~\ref{Fig1}(b-e), one must either treat the fermion self-energy and the vertex corrections on the same footing within a fully renormalized framework, or equivalently perform the calculation perturbatively using the bare fermion propagator and include the self-energy correction in Fig.~\ref{Fig1}(b) and the vertex corrections in Fig.~\ref{Fig1}(c-e) order by order.
A detailed analysis of this interplay is presented in Appendix~\ref{E}.
The scaling form used in the main text should therefore be understood within this consistent self-energy-plus-vertex treatment.

\section{Finite-temperature Optical Shear Viscosity.}
\label{4}
In the finite temperature quantum critical regime, where thermal fluctuations can be neglected, we are mainly concerned with the dc shear viscosity, which requires incorporating impurity scattering effects. This case will be discussed in the following section. 
Here, we briefly discuss the finite-temperature optical shear viscosity.
In the limit $|\Omega|\ll T$, we show in the Appendix \ref{A} that the contribution of the higher loop Feynman diagrams Fig.~\ref{Fig1}(b-e) to the shear viscosity is given by
\begin{equation}
	\begin{aligned}
	{\rm Re}[\eta(\Omega)]\approx &\frac{N^2|J|^4T^{-1}}{4N'\Omega^2}\int \frac{d^2\bm{p}}{(2\pi)^2}\int \frac{d^2\bm{k}_1}{(2\pi)^2}\int \frac{d^2\bm{k}_2}{(2\pi)^2}\int d\nu \\
		 &\times(\Delta t)^2[{\rm Im}D(i\nu\rightarrow \nu+i\eta,\bm p)]^2\\
		 &\times n_F(\epsilon_{{\bm k}_1+{\bm p}})[1-n_F(\epsilon_{{\bm k}_1})]\delta(\epsilon_{{\bm k}_1+{\bm p}}-\epsilon_{{\bm k}_1}-\nu)\\
		 &\times n_F(\epsilon_{{\bm k}_2})[1-n_F(\epsilon_{{\bm k}_2+{\bm p}})]\delta(\epsilon_{{\bm k}_2+{\bm p}}-\epsilon_{{\bm k}_2}-\nu),
	\end{aligned}
	\label{tc1}
\end{equation}
The numerator on the right-hand side coincides with the transport quantity derived from the Boltzmann-equation \cite{abrikosov2017,PhysRevLett.106.106403} and satisfies the expected $\Omega/T$ scaling relations in the quantum critical regime \cite{PhysRevB.109.115156}.
Therefore, in Appendix~\ref{C}, we infer that the finite-temperature shear viscosity is given by
$
\mathrm{Re}\big[\eta(T\gg |\Omega|)\big]\sim \frac{N^{3/2}|J|^3T^{3/2}}{N'^{1/2}\Omega^2}.
$
Combining with Eqs.\eqref{ov-T01}, we obtain the optical shear viscosity in the quantum critical regime, given by
\begin{equation}
	{\rm Re}[\eta(\Omega,T)]\sim \frac{N^{3/2}|J|^3e^\Lambda}{N'^{1/2}}\frac{|\Omega|^{3/2}+T^{3/2}}{\Omega^2}.
	\label{alaa}
\end{equation}
where a similar scaling form also appears in the optical conductivity \cite{PhysRevB.108.235125, PhysRevB.72.104510}.
In the dc limit, the shear viscosity diverges in the clean regime.

\section{dc Shear Viscosity.}
\label{5}
To obtain a finite dc shear viscosity, impurity scattering must be taken into account.
In this section, we focus on the dirty limit $\Gamma\gg |\Omega|$.
Unlike in the clean case, the one-loop diagram in Fig.~\ref{Fig1}(a) no longer vanishes after analytic continuation and gives a positive, temperature-independent background,
${\rm Re}\,[\eta_{0}(\Gamma,T,\Omega=0)]\sim N\Lambda_U^4/\Gamma>0$.
The temperature-dependent critical correction is obtained from the higher-loop diagrams in Fig.~\ref{Fig1}(b-e).
Therefore a negative sign appearing in the higher-loop dc correction should be interpreted only as a negative perturbative correction to this positive background, not as a negative total viscosity.
Within the perturbative regime considered here, the full real part of the viscosity remains nonnegative.

As demonstrated in the Appendix \ref{A}, within the Prange-Kadanoff approximation, the higher loop Feynman diagrams in Fig.~\ref{Fig1}(b-e) yield a dc shear viscosity given by
\begin{equation}
	\begin{aligned}
	&{\rm Re}[\eta(\Gamma,T,\Omega=0)]\\
	\approx &-\frac{N^2|J|^4}{16N'\Gamma^2T}\int \frac{d^2\bm{p}}{(2\pi)^2}\int \frac{d^2\bm{k}_1}{(2\pi)^2}\int \frac{d^2\bm{k}_2}{(2\pi)^2}\int d\nu \\
		 &\times(\Delta t)^2[{\rm Im}D(i\nu\rightarrow \nu+i\eta,\bm p)]^2 \\
		 &\times n_F(\epsilon_{{\bm k}_1+{\bm p}})[1-n_F(\epsilon_{{\bm k}_1})]\delta(\epsilon_{{\bm k}_1+{\bm p}}-\epsilon_{{\bm k}_1}-\nu)\\
		 &\times n_F(\epsilon_{{\bm k}_2})[1-n_F(\epsilon_{{\bm k}_2+{\bm p}})]\delta(\epsilon_{{\bm k}_2+{\bm p}}-\epsilon_{{\bm k}_2}-\nu).
	\end{aligned}
	\label{tc2}
\end{equation}
This result fully reproduces the Boltzmann-equation prediction in the impurity-dominated regime (see also Appendix~\ref{C}). 
Comparing Eqs.~\eqref{tc1} and \eqref{tc2}, we infer that the dc shear viscosity in the quantum critical regime is given by
\begin{equation}
	{\rm Re}[\eta(\Gamma,T,\Omega=0)]\sim -\frac{N^{3/2}|J|^3e^\Lambda}{N'^{1/2}}\frac{T^{3/2}}{\Gamma^2}.
	\label{dcsv1}
\end{equation}

The above inference relies \eqref{dcsv1} on the validity of the Prange-Kadanoff approximation. Unfortunately, the inclusion of impurity scattering broadens the fermion spectral function by an amount $\sim \Gamma$, so that it can no longer be regarded as exhibiting a sharp peak near the Fermi surface. As a result, in the dirty limit the boson self-energy in Eq.~\eqref{bose} no longer takes the expected Landau damping form but instead acquires an explicit dependence on the impurity scattering rate $\Gamma$. 
In order to reasonably simplify the analysis while emphasizing the role of impurities, we follow the approximation used in Ref.~\cite{science.abq6011}, in which the interaction induced fermion self-energy and the bare term $i\omega_m$ are neglected, and only the impurity scattering rate is kept. The fermion Green function then takes the form $G(i\omega_n,{\bm k})\approx (i{\rm sgn}(\omega_n)\Gamma-\epsilon_{\bm k})^{-1}$.
Applying the results in the Appendix \ref{A} and summing these contributions, we obtain
\begin{equation}
	\begin{aligned}
	\Xi(i\Omega_m)\approx &\frac{N^2|J|^4}{16N'\Gamma^2}T\sum_{\Omega}\int \frac{d^2\bm{p}}{(2\pi)^2}\int_{\Omega,{\bm p}} D[{\bm k}_1]\int_{\Omega,{\bm p}} D[{\bm k}_2]\\
		 &\times (\Delta t)^2 D(i\Omega-i\Omega_m/2,\bm{p})D(i\Omega+i\Omega_m/2,\bm{p}).
	\end{aligned}
	\label{dctc1}
\end{equation} 
In the same manner, by applying the momentum-integration formulas \eqref{GamIntk} in Appendix \ref{B}, we find that the combined contribution of the higher-loop Feynman diagrams to the dc shear viscosity reads
\begin{equation}
	\begin{aligned}
	&{\rm Re}[\eta(\Gamma,T,\Omega=0)]\\
	=&-\frac{N'\Gamma}{16\pi}\int \frac{d^2\bm{x}}{(2\pi)^2} {\bm x}^4[\frac{{\bm x}^2|\epsilon_{\bm x}|}{a f(1/|\epsilon_{\bm x}|)}{\rm \Psi}(1,\frac{{\bm x}^2|\epsilon_{\bm x}|}{a f(1/|\epsilon_{\bm x}|)})-1]\\
	\sim & N'\Gamma \int_1^{\Lambda_U^2/\Gamma} dx \int_0^\Lambda d\theta x^2 \cosh(2\theta)^2[\frac{x^2\cosh(2\theta)}{a}\\
	&\times \Psi(1,\frac{x^2\cosh(2\theta)}{a})-1]+N'\Gamma \int_0^{1} dx \int_0^\Lambda d\theta x^2 \\
	&\times \cosh(2\theta)^2[\frac{x\cosh(2\theta)}{a}\Psi(1,\frac{x\cosh(2\theta)}{a})-1]\\
	\sim &- N'\Gamma a^{3/2} \int_0^\Lambda d\theta e^{\theta} [\Lambda_U^2e^{\theta}/\Gamma\sqrt{a}-e^{\theta}/\sqrt{2a}] \\
	&+N'\Gamma a^3 \int_0^\Lambda d\theta e^{4\theta}/a^2\\
	\sim &- N|J|^2 T \frac{e^{2\Lambda} \Lambda_U^2+e^{4\Lambda}\Gamma}{\Gamma^2}
	\end{aligned}
	\label{dcsv2}
\end{equation}
where $a\equiv NT|J|^2/2N'\Gamma^2$, and $\Psi(1,b)$ is the polygamma function.
Here, $f(1/|\epsilon_{\bm x}|)$ is defined in Eq.\eqref{bose} and represents the effect of impurity scattering on the boson self-energy.
It should be understood as a model-dependent result of the
large-\(N\) random-flavor theory in the impurity-dominated regime. Its
coefficient depends explicitly on the patch cutoffs \(\Lambda_U\) and
\(\Lambda\), and is therefore not universal. The cutoff dependence originates
from the ultraviolet sensitivity of the shear-stress vertex and from the
dirty critical boson kernel. The linear-in-\(T\) behavior is the leading
temperature dependence within the model and the scale hierarchy considered
here.

The scaling in Eq.~\eqref{dcsv2} relies on the dirty boson kernel in Eq.~\eqref{bose}.
One possible concern is that the same disorder that produces the fermion scattering rate $\Gamma$ may also generate an impurity-ladder correction to the Yukawa vertex, thereby modifying the polarization kernel.
As shown explicitly in Appendix~\ref{D}, for the flavor-random matrix disorder and the random Yukawa coupling used in Eq.~\eqref{eff}, the impurity ladder contributes only to the flavor-diagonal Yukawa components.
The number of such components is $O(N)$, whereas the off-diagonal components are $O(N^2)$.
Therefore, the ladder-corrected diagonal sector is suppressed by $1/N$ relative to the bare off-diagonal sector.
The leading large-$N$ polarization kernel is thus still given by Eq.~\eqref{bose}.
Having established this, the difference between Eqs.~\eqref{dcsv1} and \eqref{dcsv2} can be traced to the singular dirty damping factor $f(1/|\epsilon_{\bm x}|)$ in the boson self-energy.
This factor changes the scaling of the integral in Eq.~\eqref{dcsv2} and leads to the dirty critical dc viscosity scaling as $T(e^{2\Lambda}\Lambda_U^2+e^{4\Lambda}\Gamma)/\Gamma^2$.
By contrast, if one artificially sets $f(1/|\epsilon_{\bm x}|)=1$ throughout, thereby replacing the dirty boson kernel by the clean Landau-damping form inside the dirty calculation, the same integral scales as $a^{3/2}$ and reproduces the Boltzmann-like result in Eq.~\eqref{dcsv1}.

Which result is ultimately more reliable? In the clean limit, the critical boson propagator acquires a broadening of order $N|J|^2|\Omega|/|\epsilon_{\bm q}|/N'\sim N|J|^2|\Omega|^{1/2}/N'$, which in the large-$N$ limit is not larger than the fermion broadening $\sqrt{N'/N}|J||\omega|^{1/2}$, and the Prange-Kadanoff approximation is therefore well justified.
By contrast, in the dirty limit the critical boson broadening, of order $|J|^2N|\Omega|/N'\Gamma$, can become much smaller than the fermion broadening $\Gamma$. The Prange-Kadanoff approximation is clearly no longer valid, since the fermion propagator cannot be approximated as a delta function sharply peaked at the Fermi surface. This breakdown ultimately leads to a linear-in-temperature dc shear viscosity.
However, away from the quantum critical regime the boson correlation length becomes finite, and the effect of the boson self-energy is subleading. In this case, the Prange-Kadanoff approximation remains valid, and the linearized Boltzmann-equation and the diagrammatic Kubo approach yield consistent results.
Remarkably, a distinct situation arises for the dc conductivity in the quantum critical regime. In this case, both the Boltzmann-equation approach and the diagrammatic Kubo formalism yield a linear-in-temperature behavior \cite{Note1}.

\section{Discussion.}
\label{6}
Throughout the above analysis, the effects of Umklapp scattering \cite{andp.19293950803} have been neglected.
However, when the VHS is located at the Brillouin zone boundary, Umklapp processes become an important scattering mechanism. 
For example, in cuprate superconductors, Umklapp scattering can connect two distinct VHS points, with ${\bm k}_1$ and ${\bm k}_2$ in Eq.\eqref{all} originating from different VHSs. In this case, the resulting shear viscosity is still consistent with the behaviors described by Eqs.\eqref{alaa} and \eqref{dcsv2}.
We then consider the shear viscosity of the entire system. 
For the convex portions of the Fermi surface away from the VHS, the leading order contribution vanishes due to $\Delta t = 0$. The nonvanishing contribution arises at next order, yielding an optical response that scales as $(\Omega^2+T^2)/\Omega^2$.

We next consider spatially random interactions superimposed on the spatially uniform interaction in the effective theory \eqref{eff} \cite{science.abq6011,PhysRevB.105.235111}. The random coupling is assumed to obey a Gaussian distribution in space. Since spatial randomness breaks momentum conservation, the resulting self-energies are momentum independent. Explicit calculations in Refs.~\cite{science.abq6011,xing2025strangemetal} further show that, owing to the reflection symmetry of the Fermi surface, only the self-energy diagram in Fig.~1(b) contributes. Combining this result with Eq.~\eqref{se}, we obtain
\begin{equation}
	\begin{aligned}
\Xi_{\mathrm{SE}}(i\Omega_m)
&\sim\Lambda_U^4\ln\frac{\Lambda_U^2+\Gamma}{|\Omega_m|}.
	\end{aligned}
\end{equation}
After analytic continuation and application of the Kubo formula, this yields interaction-induced optical conductivity and optical shear viscosity proportional to \((|\Omega|+T)/\Omega^2\) in the clean limit, for both the VHS and a generic convex Fermi surface. In the dirty limit, impurity scattering does not change the frequency-linear, momentum-independent form of the boson self-energy \cite{science.abq6011,xing2025strangemetal}. Consequently, the corresponding dc electrical conductivity and dc shear viscosity retain the scaling \(T/\Gamma^2\). Thus, unlike spatially uniform interactions, spatially random interactions produce the same scaling behavior for the VHS and a generic convex Fermi surface.

\section{Conclusion and Outlook.}
\label{7}
We have compared the diagrammatic Kubo formalism with the linearized Boltzmann-equation for shear transport near a VHS coupled to critical bosonic fluctuations.
The two approaches agree whenever the fermion spectral function remains sharply peaked and the Prange-Kadanoff reduction is valid.
In the dirty quantum-critical regime, however, our large-$N$ analysis shows that impurity broadening invalidates this sharp-fermion reduction. The leading temperature-dependent contribution to the dc shear viscosity is then controlled by the impurity-dependent bosonic damping kernel and scales as
$
T
\left(e^{2\Lambda}\Lambda_U^2+e^{4\Lambda}\Gamma\right)/\Gamma^2,$
rather than following the Boltzmann-like estimate $T^{3/2}e^\Lambda/\Gamma^2$.
These coefficients are cutoff dependent and nonuniversal. The principal qualitative result is the difference between the full dirty Kubo calculation and the sharp-fermion Boltzmann reduction within the same model. Quantitative applications to specific materials require a lattice-level description of the electronic dispersion, shear vertex, disorder, and bosonic fluctuations.

Although we have focused on a particular boson mode, Eqs.\eqref{all}, \eqref{tc2} and \eqref{dctc1}  are formulated in a way that extends straightforwardly to density wave type boson fluctuations \cite{PhysRevLett.130.126001,PhysRevLett.110.127001,PhysRevB.100.035135} (e.g., charge or spin density channels). This immediately opens a route to explore how such fluctuations reshape transport, including regimes where multiple scattering mechanisms coexist and compete.
More broadly, in realistic VHS materials, the emergence of a strange metal is often accompanied by proximity to pseudogap behavior \cite{PhysRevLett.83.3538,PhysRevLett.114.147001,doiron2017pseudogap} and superconductivity. The resulting competition can generate qualitatively new transport phenomenology beyond any single channel description. A particularly appealing direction is to leverage our framework to establish quantitative links between transport coefficients and the superconducting instability \cite{yuan2022scaling,jiang2023interplay}, for instance by correlating the strength and structure of the boson mediated scattering that governs transport with the pairing scale, thereby connecting measurable transport responses to the superconducting $T_c$.

\textit{Acknowledgments.}--
Y.-H.X. acknowledges financial support from Prof. Shang Liu funded by the Chinese
Academy of Sciences (CAS) under Grant No. YSBR-150
and a startup fund from the Institute of Physics, CAS.
W.-M. L. is supported by the National Key R\&D Program of China under Grants No.~2021YFA1400900,~2021YFA0718300, and 2021YFA1402100, NSFC under Grants No.~12174461,~12234012,~12334012, and 52327808, and the Space Application System of China Manned Space Program.

\appendix
\section{The correlation of the momentum density current}
\label{A}
Here, we derive a general expression for the contributions of all Feynman diagrams shown in Fig.~\ref{Fig1} to the momentum current correlation function.
In this appendix, we consider a boson-fermion coupled theory with arbitrary fermion dispersion $\epsilon_{\bm k}$ and form factor $f_{\bm k}$.
This type of boson-fermion coupled theory can also describe Pomeranchuk fluctuations or Coulomb interactions between electrons.
Unless otherwise stated, the clean-limit derivation uses the bare fermion Green function
$G(i\omega_n,{\bm k})=(i\omega_n-\epsilon_{\bm k})^{-1}$
and the critical boson Green function
$D^{-1}(i\Omega_m,{\bm q})=D_0^{-1}(i\Omega_m,{\bm q})-\Pi(i\Omega_m,{\bm q})$,
where $D_0^{-1}$ contains the regular mass, frequency, and momentum terms, and $\Pi$ is the boson self-energy.
The factor $f_{{\bm k}+{\bm q}/2}$ denotes the microscopic form factor of the boson-fermion vertex.
For the spatially uniform Yukawa coupling used in the main text, this form factor may be set to unity, while it is kept here to make the derivation applicable to more general density-wave or Pomeranchuk channels.

We first consider the case without impurities and begin with the self-energy diagram shown in Fig.~\ref{Fig1}(b), which reads
\begin{equation}
\begin{aligned}
	&\Xi_{\rm SE}(i\Omega_m)\\
	=&-NT\sum_n\int \frac{d^2\bm{k}}{(2\pi)^2} t_{\bm k}^2 G(i\omega_n,\bm{k})G^2(i\Omega_m+i\omega_n,\bm{k})\\
	&\times \Sigma(i\omega_n+i\Omega_m,\bm{k})+(\Omega_m\rightarrow -\Omega_m)\\
	=& \frac{N}{i\Omega_m}T\sum_n\int \frac{d^2\bm{k}}{(2\pi)^2} t_{\bm k}^2 G(i\omega_n,\bm{k})G(i\Omega_m+i\omega_n,\bm{k})\\
	&\times [\Sigma(i\omega_n,\bm{k})-\Sigma(i\omega_n+i\Omega_m,\bm{k})]\\
	=& -\frac{N}{\Omega_m^2}T\sum_n\int \frac{d^2\bm{k}}{(2\pi)^2} t_{\bm k}^2 [G(i\omega_n,\bm{k})-G(i\Omega_m+i\omega_n,\bm{k})]\\
	&\times [\Sigma(i\omega_n,\bm{k})-\Sigma(i\omega_n+i\Omega_m,\bm{k})],
	\label{se}
	\end{aligned}
\end{equation}
where 
\begin{equation}
\begin{aligned}
	\Sigma(i\omega_n,\bm{k})=&|J|^2T\sum_l\int \frac{d^2\bm{k_2}}{(2\pi)^2}G(i\omega_n+i\Omega_l,\bm{k}+\bm{k_2})\\
	&\times D(i\Omega_l,\bm{k_2})f^2_{\bm{k}+\bm{k}_2/2}	
\end{aligned}
\end{equation}
denotes the fermion self-energy.
At the second equality, we employed the identity obeyed by the bare fermion Green function, 
\begin{equation}
	G(i\omega_n+i\Omega_m,\bm{k})G(i\omega_n,\bm{k})=\frac{G(i\omega_n,\bm{k})-G(i\omega_n+i\Omega_m,\bm{k})}{i\Omega_m},
	\label{id-bare}
\end{equation}
followed by a change of integration variables in frequency, $\omega_n\rightarrow \omega_n+\Omega_m$. 
Here we neglect the self-energy induced by the Yukawa interaction, whose effect is discussed in Appendix~\ref{E}.

The vertex diagram shown in Fig.\ref{Fig1}(c) yields
\begin{equation}
\begin{aligned}
	&\Xi_{\rm V}(i\Omega_m)\\
	=&-N|J|^2T\sum_l T\sum_n\int \frac{d^2\bm{k_2}}{(2\pi)^2} \int \frac{d^2\bm{k_1}}{(2\pi)^2} t_{{\bm k}_1}t_{{\bm k}_1+{\bm k}_2} \\
	&\times G (i\omega_n,\bm{k_1})G(i\Omega_m+i\omega_n,\bm{k_1})D(i\Omega_l,\bm{k_2})f^2_{\bm{k}_1+\bm{k}_2/2}\\
	&\times G(i\omega_n+i\Omega_l+i\Omega_m,\bm{k_1}+\bm{k_2})G(i\omega_n+i\Omega_l,\bm{k_1}+\bm{k_2}).\\
	=&-\frac{N|J|^2}{\Omega_m^2}T\sum_l T\sum_n\int \frac{d^2\bm{k_2}}{(2\pi)^2} \int \frac{d^2\bm{k_1}}{(2\pi)^2} t_{{\bm k}_1}(t_{{\bm k}_1+{\bm k}_2}-t_{{\bm k}_1}) \\
	&\times [G (i\omega_n,\bm{k_1})-G(i\Omega_m+i\omega_n,\bm{k_1})]D(i\Omega_l,\bm{k_2})f^2_{\bm{k}_1+\bm{k}_2/2}\\
	&\times [G(i\omega_n+i\Omega_l+i\Omega_m,\bm{k_1}+\bm{k_2})-G(i\omega_n+i\Omega_l,\bm{k_1}+\bm{k_2})].\\
	&-\frac{N|J|^2}{\Omega_m^2}T\sum_l T\sum_n\int \frac{d^2\bm{k_2}}{(2\pi)^2} \int \frac{d^2\bm{k_1}}{(2\pi)^2} t_{{\bm k}_1}^2 \\
	&\times [G (i\omega_n,\bm{k_1})-G(i\Omega_m+i\omega_n,\bm{k_1})]D(i\Omega_l,\bm{k_2})f^2_{\bm{k}_1+\bm{k}_2/2}\\
	&\times [G(i\omega_n+i\Omega_l+i\Omega_m,\bm{k_1}+\bm{k_2})-G(i\omega_n+i\Omega_l,\bm{k_1}+\bm{k_2})].\\
	\end{aligned}
	\label{v}
\end{equation}
Notably, the second part of the vertex diagram cancels exactly with the self-energy diagram, thereby yielding a more simplified expression
\begin{equation}
	\begin{aligned}
		&\Xi_{\rm SE+V}(i\Omega_m)\\
		=&\frac{N|J|^2}{\Omega_m^2}T\sum_{\Omega_1}T\sum_{\omega}\int \frac{d^2\bm{k_2}}{(2\pi)^2} \int \frac{d^2\bm{k_1}}{(2\pi)^2} t_{{\bm k}_1}\\
		&\times (t_{{\bm k}_1+{\bm{k}_2}}-t_{{\bm k}_1}) D(i\Omega_1,\bm{k_2})f^2_{\bm{k}_1+\bm{k}_2/2}[G(i\omega,\bm{k_1})-\\
	&G(i\Omega_m+i\omega,\bm{k_1})] [G(i\Omega_1+i\omega,\bm{k_1}+\bm{k_2})-\\
	&G(i\Omega_1+i\omega+i\Omega_m,\bm{k_1}+\bm{k_2})].
	\end{aligned}
	\label{se+v1}
\end{equation}

For notational simplicity, we define the following simplified symbols:
\begin{equation}
	\begin{aligned}
		\int_{\Omega,{\bm p}} D[{\bm k}]\equiv &T\sum_{\omega}\int \frac{d^2{\bm k}}{(2\pi)^2}G(i\omega+i\Omega,\bm{k}+\bm{p})f^2_{\bm{k}+\bm{p}/2}\\
		&\times [G(i\omega-i\Omega_m/2,\bm{k})-G(i\omega+i\Omega_m/2,\bm{k})].
	\end{aligned}
	\label{id1}
\end{equation}
Then Eq.\eqref{se+v1} can be recast as
\begin{equation}
\begin{aligned}
		&\Xi_{\rm SE+V}(i\Omega_m)\\
		=&\frac{N|J|^2}{\Omega_m^2}T\sum_{\Omega}\int \frac{d^2\bm{p}}{(2\pi)^2} \int_{\Omega,{\bm p}} D[{\bm k}_1] t_{{\bm k}_1}(t_{{\bm k}_1+{\bm{p}}}-\\
		&t_{{\bm k}_1})[D(i\Omega+i\Omega_m/2,\bm{p})-D(i\Omega-i\Omega_m/2,\bm{p})].
\end{aligned}
	\label{se+v2}
\end{equation}
The boson propagator admits a similar simplification as in Eq.\eqref{id-bare}, namely,
\begin{equation}
	\begin{aligned}
		&D(i\Omega-i\Omega_m/2,\bm{p})D(i\Omega+i\Omega_m/2,\bm{p})\\
		=&\frac{D(i\Omega-i\Omega_m/2,\bm{p})-D(i\Omega+i\Omega_m/2,\bm{p})}{\Pi(i\Omega-i\Omega_m/2,\bm{p})-\Pi(i\Omega+i\Omega_m/2,\bm{p})}\\
		=&\frac{N'}{|J|^2N}\frac{D(i\Omega-i\Omega_m/2,\bm{p})-D(i\Omega+i\Omega_m/2,\bm{p})}{\int_{\Omega,{\bm p}} D[{\bm k}_2]},
	\end{aligned}
	\label{id-bos}
\end{equation}
where $\Pi$ is the boson self-energy.
Substituting this into Eq.\eqref{se+v2}, we obtain
\begin{equation}
	\begin{aligned}
		&\Xi_{\rm SE+V}(i\Omega_m)\\
		=&-\frac{N^2|J|^4}{N'\Omega_m^2}T\sum_{\Omega}\int \frac{d^2\bm{p}}{(2\pi)^2} \int_{\Omega,{\bm p}} D[{\bm k}_1]\int_{\Omega,{\bm p}} D[{\bm k}_2] \\
		&\times  t_{{\bm k}_1}(t_{{\bm k}_1+{\bm{p}}}-t_{{\bm k}_1})D(i\Omega+i\Omega_m/2,\bm{p})D(i\Omega-i\Omega_m/2,\bm{p}).
	\end{aligned}
	\label{se+v3}
\end{equation}

Finally, we evaluate the AL diagrams \cite{9789814317344_0004} depicted in Fig.~\ref{Fig1}(d-e), given by
\begin{equation}
	\begin{aligned}
		 &\Xi_{\rm AL}(i\Omega_m)\\
		 =&\Xi_{\rm AL_1}(i\Omega_m)+\Xi_{\rm AL_2}(i\Omega_m)\\
		 =&|J|^4\frac{N^2}{N'}T\sum_{\omega_1}T\sum_{\omega_2}T\sum_{\Omega}\int \frac{d^2\bm{k}_1}{(2\pi)^2}\int \frac{d^2\bm{k}_2}{(2\pi)^2}\int \frac{d^2\bm{p}}{(2\pi)^2}\\
		 &\times t_{\bm{k}_1}t_{\bm{k}_2}G(i\omega_1-i\Omega_m/2,\bm{k}_1)G(i\omega_1+i\Omega_m/2,\bm{k}_1)\\
 &\times G(i\omega_1+i\Omega,\bm{k}_1+\bm{p})f^2_{\bm{k}_1+\bm{p}/2}D(i\Omega-i\Omega_m/2,\bm{p})\\
 &\times D(i\Omega+i\Omega_m/2,\bm{p})G(i\omega_2-i\Omega_m/2,\bm{k}_2)\\
 &\times G(i\omega_2+i\Omega_m/2,\bm{k}_2)[G(i\omega_2+i\Omega,\bm{k}_2+\bm{p})f^2_{\bm{k}_2+\bm{p}/2}+\\
 &G(i\omega_2-i\Omega,\bm{k}_2-\bm{p})f^2_{\bm{k}_2-\bm{p}/2}].\\
	\end{aligned}
	\label{al1}
\end{equation}
Similarly, by repeatedly using Eq.\eqref{id-bare} for the fermion Green function and simplifying with the notation in Eq.\eqref{id1}, we obtain
\begin{equation}
	\begin{aligned}
		 &\Xi_{\rm AL}(i\Omega_m)\\
		 =&-\frac{N^2|J|^4}{N'\Omega_m^2}T\sum_{\Omega}\int \frac{d^2\bm{p}}{(2\pi)^2}\int_{\Omega,{\bm p}} D[{\bm k}_1]\{\int_{\Omega,{\bm p}} D[{\bm k}_2]+\\
		 &\int_{-\Omega,-{\bm p}} D[{\bm k}_2]\}t_{\bm{k}_1}t_{\bm{k}_2}D(i\Omega-i\Omega_m/2,\bm{p})D(i\Omega+i\Omega_m/2,\bm{p})\\
		 =&-\frac{N^2|J|^4}{N'\Omega_m^2}T\sum_{\Omega}\int \frac{d^2\bm{p}}{(2\pi)^2}\int_{\Omega,{\bm p}} D[{\bm k}_1]\int_{\Omega,{\bm p}} D[{\bm k}_2]\\
		 &\times t_{\bm{k}_1}(t_{\bm{k}_2}-t_{\bm{k}_2+{\bm p}})  D(i\Omega-i\Omega_m/2,\bm{p})D(i\Omega+i\Omega_m/2,\bm{p}).\\
	\end{aligned}
	\label{al2}
\end{equation}
At the second equality, we performed a change of variables ${\bm k}_2\rightarrow {\bm k}_2+{\bm p}$.
Adding this to the results \eqref{se+v3} from the self-energy and vertex diagrams, we obtain the leading order contribution:
\begin{equation}
	\begin{aligned}
	&\Xi(i\Omega_m)\\
	=&\Xi_{\rm SE+V}(i\Omega_m)+\Xi_{\rm AL}(i\Omega_m)\\
	=&-\frac{N^2|J|^4}{N'\Omega_m^2}T\sum_{\Omega}\int \frac{d^2\bm{p}}{(2\pi)^2}\int_{\Omega,{\bm p}} D[{\bm k}_1]\int_{\Omega,{\bm p}} D[{\bm k}_2]\\
		 &\times t_{\bm{k}_1}\Delta t  D(i\Omega-i\Omega_m/2,\bm{p})D(i\Omega+i\Omega_m/2,\bm{p}),
	\end{aligned}
	\label{all1}
\end{equation}
where $\Delta t=t_{\bm{k}_2}-t_{\bm{k}_2+{\bm p}}+t_{{\bm k}_1+{\bm{p}}}-t_{{\bm k}_1}$ represents the difference between the deformation potentials before and after scattering.
Noting that ${\bm k}_1$, ${\bm k}_2$ are internal integration variables, the labels can be interchanged. Consequently, Eq.\eqref{all1} can be rewritten as
\begin{equation}
	\begin{aligned}
	\Xi(i\Omega_m)=&\frac{N^2|J|^4}{N'\Omega_m^2}T\sum_{\Omega}\int \frac{d^2\bm{p}}{(2\pi)^2}\int_{\Omega,{\bm p}} D[{\bm k}_1]\int_{\Omega,{\bm p}} D[{\bm k}_2]\\
		 &\times t_{\bm{k}_2}\Delta t  D(i\Omega-i\Omega_m/2,\bm{p})D(i\Omega+i\Omega_m/2,\bm{p}),
	\end{aligned}
	\label{all2}
\end{equation}
We then make use of the properties of the boson propagator. 
In our case, the boson propagator is an even function of both frequency and momentum, corresponding to $D(-i\Omega,-{\bm p})=D(i\Omega,{\bm p})$. Noting further that $\int_{\Omega,{\bm p}} D[{\bm k}+\bm p]=-\int_{-\Omega,-{\bm p}} D[{\bm k}]$, we perform the change of variables $(\Omega,{\bm p})\rightarrow -(\Omega,{\bm p})$ in Eq.\eqref{all1}, followed by a shift of the momentum variable ${\bm k}_1\rightarrow {\bm k}_1+{\bm p}$ and ${\bm k}_2\rightarrow {\bm k}_2+{\bm p}$, which yields the desired expression
\begin{equation}
	\begin{aligned}
	\Xi(i\Omega_m)=&\frac{N^2|J|^4}{N'\Omega_m^2}T\sum_{\Omega}\int \frac{d^2\bm{p}}{(2\pi)^2}\int_{\Omega,{\bm p}} D[{\bm k}_1]\int_{\Omega,{\bm p}} D[{\bm k}_2]\\
		 &\times t_{\bm{k}_1+{\bm p}}\Delta t D(i\Omega-i\Omega_m/2,\bm{p})D(i\Omega+i\Omega_m/2,\bm{p}).
	\end{aligned}
	\label{all3}
\end{equation}
The same transformations can also be applied to Eq.\eqref{all2} and obtain
\begin{equation}
	\begin{aligned}
	\Xi(i\Omega_m)=&-\frac{N^2|J|^4}{N'\Omega_m^2}T\sum_{\Omega}\int \frac{d^2\bm{p}}{(2\pi)^2}\int_{\Omega,{\bm p}} D[{\bm k}_1]\int_{\Omega,{\bm p}} D[{\bm k}_2]\\
		 &\times t_{\bm{k}_2+{\bm p}}\Delta t D(i\Omega-i\Omega_m/2,\bm{p})D(i\Omega+i\Omega_m/2,\bm{p}).
	\end{aligned}
	\label{all33}
\end{equation}
Summing the resulting expressions with Eqs.\eqref{all1}, \eqref{all2}, \eqref{all3} and \eqref{all33}, we finally obtain
\begin{equation}
	\begin{aligned}
	\Xi(i\Omega_m)=&\frac{N^2|J|^4}{4N'\Omega_m^2}T\sum_{\Omega}\int \frac{d^2\bm{p}}{(2\pi)^2}\int_{\Omega,{\bm p}} D[{\bm k}_1]\int_{\Omega,{\bm p}} D[{\bm k}_2]\\
		 &(\Delta t)^2 D(i\Omega-i\Omega_m/2,\bm{p})D(i\Omega+i\Omega_m/2,\bm{p}),
	\end{aligned}
	\label{all4}
\end{equation}
This expression corresponds to Eq.\eqref{all} in the main text and serves as the starting point for all calculations of the shear viscosity in the clean limit.
Our main point is that the leading-order expression for the momentum current correlation function in Eq.\eqref{all4} is universal, it holds at both zero and finite temperatures and is valid for arbitrary forms of the fermion dispersion. 
Our derivation relies solely on the assumption that the boson Green function is even in frequency and momentum, which holds in systems preserving time-reversal symmetry.

In the finite temperature regime $|\Omega|\ll T$, Eq.\eqref{all4} can be further simplified. We perform the Matsubara-frequency sums in Eq.\eqref{all4}. Although this yields a lengthy expression, we are only interested in ${\rm Im}\Xi(i\Omega_m\rightarrow \Omega+i\eta)$. We expand the result in powers of the external frequency $\Omega$. For a time-reversal invariant system, the even powers of $\Omega$ cancel, so only odd powers contribute.
Expanding to linear order in $\Omega$ and neglecting the Drude term [proportional to $\delta(\Omega)$], we obtain the optical shear viscosity as
\begin{equation}
	\begin{aligned}
	&{\rm Re}[\eta(\Omega)]\\
	\approx &\frac{N^2|J|^4}{4N'\Omega_m^2T}\int \frac{d^2\bm{p}}{(2\pi)^2}\int \frac{d^2\bm{k}_1}{(2\pi)^2}\int \frac{d^2\bm{k}_2}{(2\pi)^2}\int d\nu \\
		 &\times(\Delta t)^2[{\rm Im}D(i\nu\rightarrow \nu+i\eta,\bm p)]^2 n_B^2(\nu)e^{\nu/T}\\
		 &\times f^2_{\bm{k}_1+\bm{p}/2}[n_F(\epsilon_{{\bm k}_1+{\bm p}})-n_F(\epsilon_{{\bm k}_1})]\delta(\epsilon_{{\bm k}_1+{\bm p}}-\epsilon_{{\bm k}_1}-\nu)\\
		 &\times f^2_{\bm{k}_2+\bm{p}/2}[n_F(\epsilon_{{\bm k}_2+{\bm p}})-n_F(\epsilon_{{\bm k}_2})]\delta(\epsilon_{{\bm k}_2+{\bm p}}-\epsilon_{{\bm k}_2}-\nu),
	\end{aligned}
	\label{all5}
\end{equation}
where $n_B$ ($n_F$) denotes the Bose (Fermi) distribution function, and $n_B^2(\nu)e^{\nu/T}$ arises from the Taylor expansion of the Bose distribution. Noting that $[n_F(a)-n_F(b)]n_B(a-b)=-n_F(a)[1-n_F(b)]$ and $[n_F(a)-n_F(b)]n_B(a-b)e^{a-b}=-n_F(b)[1-n_F(a)]$, the above expression can be further simplified to
\begin{equation}
	\begin{aligned}
	{\rm Re}[\eta(\Omega)]\approx &\frac{N^2|J|^4}{4N'\Omega^2T}\int \frac{d^2\bm{p}}{(2\pi)^2}\int \frac{d^2\bm{k}_1}{(2\pi)^2}\int \frac{d^2\bm{k}_2}{(2\pi)^2}\int d\nu \\
		 &\times(\Delta t)^2[{\rm Im}D(i\nu\rightarrow \nu+i\eta)]^2 f^2_{\bm{k}_1+\bm{p}/2}f^2_{\bm{k}_2+\bm{p}/2}\\
		 &\times n_F(\epsilon_{{\bm k}_1+{\bm p}})[1-n_F(\epsilon_{{\bm k}_1})]\delta(\epsilon_{{\bm k}_1+{\bm p}}-\epsilon_{{\bm k}_1}-\nu)\\
		 &\times n_F(\epsilon_{{\bm k}_2})[1-n_F(\epsilon_{{\bm k}_2+{\bm p}})]\delta(\epsilon_{{\bm k}_2+{\bm p}}-\epsilon_{{\bm k}_2}-\nu),
	\end{aligned}
	\label{all6}
\end{equation}
The numerator on the right-hand side of the equality coincides with the result obtained from the Boltzmann-equation. This establishes, at a preliminary level, the equivalence between the Boltzmann-equation approach and the Feynman diagrammatic formulation in the clean limit.

Finally, we discuss the effects induced by impurities.
Turning on impurity scattering, we assume that the system is characterized by an impurity scattering rate $\Gamma$. The fermion Green function is then renormalized to $G^{-1}(i\omega_n,{\bm k})=i\omega_n+i{\rm sgn}(\omega_n)\Gamma-\epsilon_{\bm k}-\Sigma(i\omega_n)$,
where $\Sigma(i\omega_n)$ denotes the self-energy arising from interactions other than impurity scattering. 
In the dirty limit $|\omega|\ll \Gamma$, the interaction-induced fermion self-energy is much smaller than the impurity scattering rate $\Gamma$, leading to
\begin{equation}
	G(i\omega_n,{\bm k})\approx \frac{1}{i\omega_n+i{\rm sgn}(\omega_n)\Gamma-\epsilon_{\bm k}}.
\end{equation}
The only difference from the clean limit $|\omega|\gg \Gamma$ case discussed above is that Eq.\eqref{id-bare} is modified to
\begin{equation}
\begin{aligned}
	&G(i\omega_n+i\Omega_m,\bm{k})G(i\omega_n,\bm{k})\\
	\approx &\frac{G(i\omega_n,\bm{k})-G(i\omega_n+i\Omega_m,\bm{k})}{i\Omega_m+i\Gamma[{\rm sgn}(\omega_n+\Omega_m)-{\rm sgn}(\omega_n)]}.
	\end{aligned}
	\label{id-dir}
\end{equation}

For the fermion Green functions in the numerator, we continue to employ the Prange-Kadanoff approximation, i.e., 
$G^R(\omega,\bm{k})\sim \delta(\omega-\epsilon_{k})$, 
Eq.\eqref{id-dir} can be further simplified to
\begin{equation}
\begin{aligned}
	G(i\omega_n+i\Omega_m,\bm{k})G(i\omega_n,\bm{k})
	\approx &\frac{G(i\omega_n,\bm{k})-G(i\omega_n+i\Omega_m,\bm{k})}{i\Omega_m+2i\Gamma{\rm sgn}(\Omega_m)}.
\end{aligned}
\end{equation}
Therefore, all steps performed in the impurity-free case remain valid in the dirty limit, except that the frequency $|\Omega|$ in the denominator is replaced by the impurity scattering rate $\Gamma$. The final result is thus given by
\begin{equation}
	\begin{aligned}
	\Xi(i\Omega_m)\approx &\frac{N^2|J|^4}{4N'(|\Omega_m|+2\Gamma)^2}T\sum_{\Omega}\int \frac{d^2\bm{p}}{(2\pi)^2}\int_{\Omega,{\bm p}} D[{\bm k}_1]\\
		 &\times \int_{\Omega,{\bm p}} D[{\bm k}_2](\Delta t)^2 D(i\Omega-i\Omega_m/2,\bm{p})\\
		 &\times D(i\Omega+i\Omega_m/2,\bm{p}),
	\end{aligned}
	\label{dir1}
\end{equation}
and
\begin{equation}
	\begin{aligned}
	&{\rm Re}[\eta(T,\Omega=0)]\\
	\approx &-\frac{N^2|J|^4}{16N'\Gamma^2T}\int \frac{d^2\bm{p}}{(2\pi)^2}\int \frac{d^2\bm{k}_1}{(2\pi)^2}\int \frac{d^2\bm{k}_2}{(2\pi)^2}\int d\nu \\
		 &\times(\Delta t)^2[{\rm Im}D(i\nu\rightarrow \nu+i\eta,\bm p)]^2 f^2_{\bm{k}_1+\bm{p}/2}f^2_{\bm{k}_2+\bm{p}/2}\\
		 &\times n_F(\epsilon_{{\bm k}_1+{\bm p}})[1-n_F(\epsilon_{{\bm k}_1})]\delta(\epsilon_{{\bm k}_1+{\bm p}}-\epsilon_{{\bm k}_1}-\nu)\\
		 &\times n_F(\epsilon_{{\bm k}_2})[1-n_F(\epsilon_{{\bm k}_2+{\bm p}})]\delta(\epsilon_{{\bm k}_2+{\bm p}}-\epsilon_{{\bm k}_2}-\nu).
	\end{aligned}
	\label{dir2}
\end{equation}
In the regime where the impurity scattering rate dominates over the interaction induced scattering rate, Eq.\eqref{dir2} reproduces transport results that are in complete agreement with the Boltzmann-equation approach \cite{PhysRevLett.106.106403}.

\section{Approximation Schemes in Different Limits}
\label{B}
Here, we present the method used to evaluate the momentum integrals in the calculation of the shear viscosity.

In the clean limit at zero temperature, the low-frequency approximation can always be applied, namely, $G(i\omega+i\Omega,{\bm k})-G(i\omega-i\Omega,{\bm k})\approx -i\pi[{\rm sgn}(\omega+\Omega)-{\rm sgn}(\omega-\Omega)]\delta(\epsilon_{\bm k})$. 
Therefore, the following formulas, 
\begin{equation}
	\begin{aligned}
		&\int \frac{d^2\bm{k}}{4\pi^2}\delta(\epsilon_{\bm k})\delta(\epsilon_{{\bm k}+{\bm p}})=\frac{1}{4\pi^2|\epsilon_{\bm q}|},\\
		&\int \frac{d^2\bm{k}}{4\pi^2}k_x\delta(\epsilon_{\bm k})\delta(\epsilon_{{\bm k}+{\bm p}})=-\frac{p_x}{8\pi^2|\epsilon_{\bm q}|},\\
		&\int \frac{d^2\bm{k}}{4\pi^2}k_x^2\delta(\epsilon_{\bm k})\delta(\epsilon_{{\bm k}+{\bm p}})=\frac{{\bm p}^2}{16\pi^2|\epsilon_{\bm q}|},\\
		&\int \frac{d^2\bm{k}}{4\pi^2}k_xk_y\delta(\epsilon_{\bm k})\delta(\epsilon_{{\bm k}+{\bm p}})=\frac{p_xp_y}{8\pi^2|\epsilon_{\bm q}|},\\		
	\end{aligned}
	\label{delintk}
\end{equation}
can always be used to simplify the calculation in Eq.\eqref{all}.

In the finite-temperature dirty limit, the fermion Green function can be approximated as $G(i\omega+i\Omega,{\bm k})-G(i\omega-i\Omega,{\bm k})\approx -i[{\rm sgn}(\omega+\Omega)-{\rm sgn}(\omega-\Omega)]\Gamma/(\Gamma^2+\epsilon_{\bm k})$. Consequently, the following relations,
\begin{equation}
	\begin{aligned}
		&\int \frac{d^2\bm{k}}{4\pi^2}\frac{\Gamma/\pi}{\epsilon_{\bm k}^2+\Gamma^2}\frac{\Gamma/\pi}{\epsilon_{\bm{k}+\bm{q}}^2+\Gamma^2}=\frac{1}{4\pi^2|\epsilon_{\bm q}|} f(\frac{\Gamma}{|\epsilon_{\bm q}|}),\\
		&\int \frac{d^2\bm{k}}{4\pi^2}k_x\frac{\Gamma/\pi}{\epsilon_{\bm k}^2+\Gamma^2}\frac{\Gamma/\pi}{\epsilon_{\bm{k}+\bm{q}}^2+\Gamma^2}=-\frac{p_x}{8\pi^2|\epsilon_{\bm p}|} f(\frac{\Gamma}{|\epsilon_{\bm p}|}),\\
		&\int \frac{d^2\bm{k}}{4\pi^2}k_x^2\frac{\Gamma/\pi}{\epsilon_{\bm k}^2+\Gamma^2}\frac{\Gamma/\pi}{\epsilon_{\bm{k}+\bm{q}}^2+\Gamma^2}=\frac{{\bm p}^2}{16\pi^2|\epsilon_{\bm p}|} f(\frac{\Gamma}{|\epsilon_{\bm p}|}),\\
		&\int \frac{d^2\bm{k}}{4\pi^2}k_xk_y\frac{\Gamma/\pi}{\epsilon_{\bm k}^2+\Gamma^2}\frac{\Gamma/\pi}{\epsilon_{\bm{k}+\bm{q}}^2+\Gamma^2}=\frac{p_xp_y}{8\pi^2|\epsilon_{\bm p}|} f(\frac{\Gamma}{|\epsilon_{\bm p}|}),\\		
	\end{aligned}
	\label{GamIntk}
\end{equation}
can be used to simplify the evaluation of Eq.\eqref{dctc1}. Note that
\begin{equation}
\begin{aligned}
	&f(x)\\
	=&\frac{x}{\pi}\int dy\frac{\mathrm{Abs}(1+y)}{y^2(1+y)^2+x^2\{1+2y(1+y)+2\mathrm{Abs}[y(1+y)]\}}
\end{aligned}
\end{equation}
also appears in the boson self-energy in Eq.\eqref{bose}.

\begin{figure}
		\centering
		\includegraphics[width=\columnwidth]{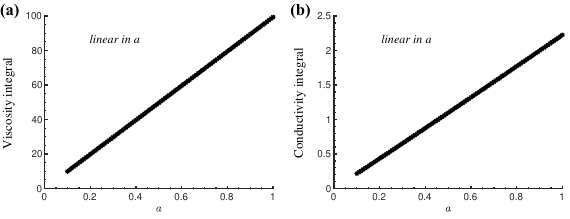}
		\caption{Numerical results for the integral in (a) Eq.\eqref{a21} and (b) Eq.\eqref{a22} as a function of $a$.
		}
		\label{Fig2}
	\end{figure}

For the boson momentum integral in Eq.\eqref{dctc1}, denoted by
\begin{equation}
	\int \frac{d^2\bm{x}}{(2\pi)^2} {\bm x}^4[\frac{{\bm x}^2|\epsilon_{\bm x}|}{a f(1/|\epsilon_{\bm x}|)}{\rm \Psi}(1,\frac{{\bm x}^2|\epsilon_{\bm x}|}{a f(1/|\epsilon_{\bm x}|)})-1],
	\label{a21}
\end{equation}
the numerical result is shown in Fig.~\ref{Fig2}(a), where we find a linear dependence on a. If the function $f(1/|\epsilon_{\bm x}|)$ is replaced by unity, Eq.\eqref{a21} can be simplified by a change of variables $x/a^{1/4}\rightarrow x$, yielding an integral that is likewise proportional to $a^{3/2}$.

Similarly, for the dc conductivity at the VHS, one encounters an analogous momentum integral, denoted by
\begin{equation}
	\int \frac{d^2\bm{x}}{(2\pi)^2} p_x^2[\frac{{\bm x}^2|\epsilon_{\bm x}|}{a f(1/|\epsilon_{\bm x}|)}{\rm \Psi}(1,\frac{{\bm x}^2|\epsilon_{\bm x}|}{a f(1/|\epsilon_{\bm x}|)})-1].
	\label{a22}
\end{equation}
As shown in Fig.~\ref{Fig2}(b), the numerical evaluation yields a linear dependence on $a$, in agreement with the result obtained when $f(1/|\epsilon_{\bm x}|)$ is replaced by unity.

\section{Relation to the Boltzmann-equation}
\label{C}
In this appendix, we clarify the relation between the diagrammatic Kubo expression and the Boltzmann result.
The finite-temperature expression in Eq.~\eqref{all6} has the structure
\begin{equation}
\begin{aligned}
{\cal I}_{\eta}
=&
T^{-1}\int_{\bm p,{\bm k}_1,{\bm k}_2,\nu}
(\Delta t)^2
[{\rm Im}D^R(\nu,{\bm p})]^2
f^2_{{\bm k}_1+{\bm p}/2}
f^2_{{\bm k}_2+{\bm p}/2}
\\
&\times
n_F(\epsilon_{{\bm k}_1+{\bm p}})
[1-n_F(\epsilon_{{\bm k}_1})]
\delta(\epsilon_{{\bm k}_1+{\bm p}}-\epsilon_{{\bm k}_1}-\nu)
\\
&\times
n_F(\epsilon_{{\bm k}_2})
[1-n_F(\epsilon_{{\bm k}_2+{\bm p}})]
\delta(\epsilon_{{\bm k}_2+{\bm p}}-\epsilon_{{\bm k}_2}-\nu).
\end{aligned}
\end{equation}
This can be deduced from the collision integral obtained from the linearized Boltzmann equation for scattering mediated by a boson with spectral weight $\mathrm{Im}D^R$ \cite{PhysRevLett.106.106403}.
The factor $\Delta t$ is the change of the stress or deformation-potential vertex during a two-particle scattering event.
Therefore, whenever the fermion spectral function can be approximated by a sharp peak at the Fermi surface, the diagrammatic expression reduces to the Boltzmann or Prange-Kadanoff result.
In the clean quantum-critical regime this gives the scaling
\begin{equation}
{\rm Re}\,\eta(\Omega,T)\sim \frac{{\cal I}_{\eta}}{\Omega^2}\sim \frac{|\Omega|^{3/2}}{\Omega^2} (T=0)\overset{\rm Scaling}{\rightarrow }\frac{T^{3/2}}{\Omega^2}
\quad (T\gg |\Omega|),
\end{equation}
this yields Eq.~\eqref{alaa} in the main text.

For the case of finite impurities, if one continues to employ the Prange-Kadanoff approximation, the shear viscosity obtained from the diagrammatic Feynman approach is identical to that from the Boltzmann equation. 
Its relation to ${\cal I}_{\eta}$ is given by
\begin{equation}
	{\rm Re}\,\eta(\Omega,T)\sim \frac{{\cal I}_{\eta}}{\Gamma^2}\sim \frac{T^{3/2}}{\Gamma^2},
\end{equation}
which reproduces Eq.\eqref{dcsv1} in the main text. 
However, in the dirty critical regime, the fermion spectral function is broadened by the impurity scattering rate $\Gamma$ and can no longer be approximated by a delta function. 
As a result, the sharp-fermion reduction fails, and one must instead retain the full diagrammatic expression with the dirty boson kernel. This yields Eq.~\eqref{dcsv2}.

\section{Impurity Vertex Ladder and Large-$N$ Flavor Counting}
\label{D}
Here we show why the impurity ladder correction to the Yukawa vertex is subleading in the large-$N$ limit for the model in Eq.~\eqref{eff}.
The disorder potential is taken to be a flavor-random matrix with
\begin{equation}
\overline{w_{mn}({\bm x})w^{*}_{m'n'}({\bm x}')}
=
|w|^2\delta_{mm'}\delta_{nn'}\delta({\bm x}-{\bm x}').
\end{equation}
The Yukawa bilinear coupled to the boson is
\begin{equation}
O_l({\bm q})
=
\sum_{{\bm k},mn}
J^{(l)}_{mn}
\psi^\dagger_{{\bm k}+{\bm q},m}
\psi_{{\bm k},n}.
\end{equation}
A representative impurity correction to this vertex contains the flavor structure
\begin{equation}
J_{mn}w_{ls}w^{*}_{qp}
\psi_m^\dagger\psi_n
\psi_l^\dagger\psi_s
\psi_p^\dagger\psi_q .
\end{equation}
Contracting the internal fermion lines around the vertex enforces the flavor indices in such a way that the disorder factors reduce to
\begin{equation}
J_{mn}w_{nn}w^{*}_{mm}.
\end{equation}
After disorder averaging,
\begin{equation}
\overline{w_{nn}w^{*}_{mm}}=|w|^2\delta_{mn}.
\end{equation}
Thus an impurity ladder correction to the Yukawa vertex exists only for the flavor-diagonal components $J_{mm}$.
The off-diagonal components $J_{mn}$ with $m\neq n$ do not acquire this ladder correction.
Since there are only $O(N)$ diagonal components but $O(N^2)$ off-diagonal components, the ladder-corrected sector is suppressed by $1/N$ in the large-$N$ expansion.
Consequently, the leading boson self-energy is controlled by the bare off-diagonal Yukawa vertices and scales as $N/N'$.
The conventional diffusion pole associated with a conserved charge density is tied to the flavor-singlet diagonal channel and is not the leading response of the random Yukawa bilinear considered here.
Therefore the dirty polarization kernel in Eq.~\eqref{bose} remains valid at leading order in $1/N$.

As a consistency check, we verify this point explicitly in the $N=1$ limit. 
In this case, the impurity ladder correction to the Yukawa vertex is nonzero. 
Using the impurity-dressed Yukawa vertex, the corresponding boson self-energy takes the form
\begin{equation}
	\begin{aligned}
		&\Pi_J(i\Omega,\bm p)
=
-|J|^2\int_\omega
\frac{
B(i\Omega,\bm p)
}{
1-|w|^2B(i\Omega,\bm p)
},\\
&B(i\Omega,\bm p)
=
\int_{\bm k}
G(i\omega+i\Omega,\bm k+\bm p)G(i\omega,\bm k).
	\end{aligned}
\end{equation}
In the zero-momentum limit near the VHS, $B$ is enhanced by the divergent density of states. Therefore, for $|w|^2B\gg1$,
\begin{equation}
\Pi_J(i\Omega,0)
\sim
-\frac{|J|^2\Lambda_\omega}{|w|^2}.
\end{equation}
This term is cutoff dependent and frequency independent, and therefore should be absorbed into the renormalized boson mass. It does not violate the Ward identity.

\section{Dressed Fermions and Current-Vertex Corrections}
\label{E}
We finally comment on the use of dressed fermion propagators.
For bare fermions,
\begin{equation}
G(i\omega+i\Omega,{\bm k})G(i\omega,{\bm k})
=
\frac{
G(i\omega,{\bm k})-G(i\omega+i\Omega,{\bm k})
}{i\Omega}.
\end{equation}
At the quantum critical point, however, the fermion self-energy behaves as
$\Sigma(\omega)\sim i\,{\rm sgn}(\omega)|\omega|^{1/2}$, so a product of two dressed Green functions contains the denominator
\begin{equation}
i\Omega-\Sigma(\omega+\Omega)+\Sigma(\omega),
\end{equation}
rather than simply $i\Omega$.
Therefore, if dressed fermion propagators are used, the corresponding vertex correction must also be included.
For a density vertex, the Ward identity gives
\begin{equation}
G(\omega+\Omega)\Gamma(\omega+\Omega,\omega)G(\omega)
=
\frac{G(\omega)-G(\omega+\Omega)}{i\Omega},
\end{equation}
with
\begin{equation}
\Gamma(\omega+\Omega,\omega)
=
1-
\frac{\Sigma(\omega+\Omega)-\Sigma(\omega)}{i\Omega}
\simeq
1-\frac{\partial\Sigma(\omega)}{\partial i\omega}.
\end{equation}
For the fermion self-energy in a non-Fermi-liquid form, at low frequencies, $\frac{\partial \Sigma(\omega)}{\partial i\omega}$ diverges, so it gives the dominant contribution.
The stress vertex considered in this work contains the additional momentum-dependent form factor $t_{\bm k}=k_xk_y$, so it is not fixed by the density Ward identity alone.
The full vertex satisfies a Bethe--Salpeter equation,
\begin{equation}
\begin{aligned}
V_J({\bm k};\omega+\Omega,\omega)
=&\,
t_{\bm k}
+
|J|^2
\int_{\omega',{\bm k}'}
D({\bm k}-{\bm k}',\omega-\omega')
\\
&\times
G({\bm k}',\omega'+\Omega)
G({\bm k}',\omega')
V_J({\bm k}';\omega'+\Omega,\omega').
\end{aligned}
\end{equation}
If the fermion self-energy is momentum independent, the leading low-energy correction near the VHS projects back into the same symmetry channel as $t_{\bm k}$.
Indeed, at ${\bm k}=0$ the correction vanishes by symmetry, because $t_{\bm k}=k_xk_y$ is odd under either $k_x\to -k_x$ or $k_y\to -k_y$.
The leading nonvanishing term is therefore proportional to $k_xk_y$.
Thus one may write asymptotically
\begin{equation}
V_J({\bm k};\omega+\Omega,\omega)
=
t_{\bm k}
\left[
1
-
A\frac{\partial\Sigma(\omega)}{\partial i\omega}
\right]
+\text{less singular terms},
\end{equation}
where $A$ is a nonuniversal constant determined by the VHS patch and the interaction kernel.
This shows that the leading singular structure of the dressed vertex compensates the corresponding singularity in the dressed fermion product, as expected in a consistent conserving calculation.
A fully self-consistent treatment with momentum-dependent self-energy is beyond the scope of the present work, but it does not affect the leading perturbative large-$N/N'$ results derived in the main text.

\bibliographystyle{elsarticle-num}
\bibliography{paper.bib}

@article{PhysRev.134.A566,
  author  = {Prange, Richard E. and Kadanoff, Leo P.},
  title   = {Transport Theory for Electron-Phonon Interactions in Metals},
  journal = {Phys. Rev.},
  volume  = {134},
  pages   = {A566--A580},
  year    = {1964},
  doi     = {10.1103/PhysRev.134.A566},
  url     = {https://doi.org/10.1103/PhysRev.134.A566}
}

@article{PhysRevB.109.115156,
  author  = {Gindikin, Yasha and Chubukov, Andrey V.},
  title   = {{Fermi} surface geometry and optical conductivity of a two-dimensional electron gas near an {Ising}-nematic quantum critical point},
  journal = {Phys. Rev. B},
  volume  = {109},
  pages   = {115156},
  year    = {2024},
  doi     = {10.1103/PhysRevB.109.115156},
  url     = {https://doi.org/10.1103/PhysRevB.109.115156}
}

@article{xing2025strangemetal,
  author  = {Xing, Yi-Hui and Liu, Wu-Ming and Zhang, Xiao-Tian},
  title   = {Strange metal at the {Lifshitz} transition},
  journal = {Sci. China Phys. Mech. Astron.},
  volume  = {69},
  pages   = {127414},
  year    = {2026},
  doi     = {10.1007/s11433-026-3106-6},
  url     = {https://doi.org/10.1007/s11433-026-3106-6}
}

@article{science.abq6011,
  author  = {Aavishkar A. Patel  and Haoyu Guo  and Ilya Esterlis  and Subir Sachdev},
  title   = {Universal theory of strange metals from spatially random interactions},
  journal = {Science},
  volume  = {381},
  pages   = {790--793},
  year    = {2023},
  doi     = {10.1126/science.abq6011},
  url     = {https://doi.org/10.1126/science.abq6011}
}

@article{PhysRevB.92.054305,
  author  = {Allen, Philip B.},
  title   = {Electron self-energy and generalized {Drude} formula for infrared conductivity of metals},
  journal = {Phys. Rev. B},
  volume  = {92},
  pages   = {054305},
  year    = {2015},
  doi     = {10.1103/PhysRevB.92.054305},
  url     = {https://doi.org/10.1103/PhysRevB.92.054305}
}

@article{PhysRevB.108.235125,
  author  = {Li, Songci and Sharma, Prachi and Levchenko, Alex and Maslov, Dmitrii L.},
  title   = {Optical conductivity of a metal near an {Ising}-nematic quantum critical point},
  journal = {Phys. Rev. B},
  volume  = {108},
  pages   = {235125},
  year    = {2023},
  doi     = {10.1103/PhysRevB.108.235125},
  url     = {https://doi.org/10.1103/PhysRevB.108.235125}
}

@article{PhysRevB.72.104510,
  author  = {Rosch, A. and Howell, P. C.},
  title   = {Zero-temperature optical conductivity of ultraclean {Fermi} liquids and superconductors},
  journal = {Phys. Rev. B},
  volume  = {72},
  pages   = {104510},
  year    = {2005},
  doi     = {10.1103/PhysRevB.72.104510},
  url     = {https://doi.org/10.1103/PhysRevB.72.104510}
}

@article{PhysRevLett.106.106403,
  author  = {Maslov, Dmitrii L. and Yudson, Vladimir I. and Chubukov, Andrey V.},
  title   = {Resistivity of a non-{Galilean}-invariant {Fermi} liquid near {Pomeranchuk} quantum criticality},
  journal = {Phys. Rev. Lett.},
  volume  = {106},
  pages   = {106403},
  year    = {2011},
  doi     = {10.1103/PhysRevLett.106.106403},
  url     = {https://doi.org/10.1103/PhysRevLett.106.106403}
}

@misc{Note1,
  note = {For the conductivity in the additional channel, one needs to consider the current--current correlation function, in which the deformation potential $t_{\bm k}$ is replaced by the group velocity $v_{\bm k}$. At the VHS, the group velocity is linear in momentum. As a result, the dc conductivity can be obtained by directly replacing the $\bm p^{4}$ factor in the numerator of Eq.~\eqref{dcsv2} with $p_x^{2}$. Remarkably, this procedure yields a linear-in-temperature dc conductivity regardless of whether the system is in the dirty regime, i.e., independently of whether one sets $f(\Gamma/|\epsilon_{\bm k}|) \equiv 1$.}
}

@article{PhysRevB.105.235111,
  author  = {Aldape, Erik E. and Cookmeyer, Tessa and Patel, Aavishkar A. and Altman, Ehud},
  title   = {Solvable theory of a strange metal at the breakdown of a heavy {Fermi} liquid},
  journal = {Phys. Rev. B},
  volume  = {105},
  pages   = {235111},
  year    = {2022},
  doi     = {10.1103/PhysRevB.105.235111},
  url     = {https://doi.org/10.1103/PhysRevB.105.235111}
}

@article{andp.19293950803,
  author  = {Peierls, Rudolf},
  title   = {Zur kinetischen Theorie der Wärmeleitung in Kristallen},
  journal = {Ann. Phys. (Berlin)},
  volume  = {395},
  pages   = {1055--1101},
  year    = {1929},
  doi     = {10.1002/andp.19293950803},
  url     = {https://doi.org/10.1002/andp.19293950803}
}

@book{abrikosov2017,
  author    = {Abrikosov, A. A.},
  title     = {Fundamentals of the Theory of Metals},
  publisher = {Dover Publications},
  address   = {Mineola, NY},
  year      = {2017},
  isbn      = {978-0-486-81901-3},
  url       = {https://store.doverpublications.com/products/9780486825762}
}

@incollection{9789814317344_0004,
  author    = {Aslamazov, L. G. and Larkin, A. I.},
  title     = {Effect of fluctuations on the properties of a superconductor above the critical temperature},
  booktitle = {30 Years of the Landau Institute---Selected Papers},
  editor    = {Khalatnikov, Isaak M. and Mineev, Vladimir P.},
  publisher = {World Scientific},
  address   = {Singapore},
  pages     = {23--28},
  year      = {1996},
  doi       = {10.1142/9789814317344_0004},
  url       = {https://doi.org/10.1142/9789814317344_0004}
}

@article{RevModPhys.58.323,
  author  = {Rammer, J. and Smith, H.},
  title   = {Quantum field-theoretical methods in transport theory of metals},
  journal = {Rev. Mod. Phys.},
  volume  = {58},
  pages   = {323--359},
  year    = {1986},
  doi     = {10.1103/RevModPhys.58.323},
  url     = {https://doi.org/10.1103/RevModPhys.58.323}
}

@article{Maslov_2017,
  author  = {Maslov, Dmitrii L and Chubukov, Andrey V},
  title   = {Optical response of correlated electron systems},
  journal = {Rep. Prog. Phys.},
  volume  = {80},
  pages   = {026503},
  year    = {2017},
  doi     = {10.1088/1361-6633/80/2/026503},
  url     = {https://doi.org/10.1088/1361-6633/80/2/026503}
}

@article{HOLSTEIN1964410,
  author  = {T. Holstein},
  title   = {Theory of transport phenomena in an electron-phonon gas},
  journal = {Ann. Phys. (N.Y.)},
  volume  = {29},
  pages   = {410--535},
  year    = {1964},
  doi     = {10.1016/0003-4916(64)90008-9},
  url     = {https://doi.org/10.1016/0003-4916(64)90008-9}
}

@article{PhysRevB.27.5775,
  author  = {Riseborough, Peter S.},
  title   = {Spin-fluctuation contribution to the high-frequency electrical conductivity of nearly magnetic transition metals},
  journal = {Phys. Rev. B},
  volume  = {27},
  pages   = {5775--5783},
  year    = {1983},
  doi     = {10.1103/PhysRevB.27.5775},
  url     = {https://doi.org/10.1103/PhysRevB.27.5775}
}

@article{PhysRevLett.97.096604,
  author  = {Farid, Abdel-Khalek and Mishchenko, Eugene G.},
  title   = {Optical Conductivity of a Two-Dimensional Electron Liquid with Spin-Orbit Interaction},
  journal = {Phys. Rev. Lett.},
  volume  = {97},
  pages   = {096604},
  year    = {2006},
  doi     = {10.1103/PhysRevLett.97.096604},
  url     = {https://doi.org/10.1103/PhysRevLett.97.096604}
}

@article{Cao2020,
  author  = {Cao, Yuan and Chowdhury, Debanjan and Rodan-Legrain, Daniel and Rubies-Bigorda, Oriol and Watanabe, Kenji and Taniguchi, Takashi and Senthil, T. and Jarillo-Herrero, Pablo},
  title   = {Strange metal in magic-angle graphene with near-{Planckian} dissipation},
  journal = {Phys. Rev. Lett.},
  volume  = {124},
  pages   = {076801},
  year    = {2020},
  doi     = {10.1103/PhysRevLett.124.076801},
  url     = {https://doi.org/10.1103/PhysRevLett.124.076801}
}

@article{Jaoui2022,
  author  = {Jaoui,  Alexandre and Das,  Ipsita and Di Battista,  Giorgio and Díez-Mérida,  Jaime and Lu,  Xiaobo and Watanabe,  Kenji and Taniguchi,  Takashi and Ishizuka,  Hiroaki and Levitov,  Leonid and Efetov,  Dmitri K.},
  title   = {Quantum critical behaviour in magic-angle twisted bilayer graphene},
  journal = {Nat. Phys.},
  volume  = {18},
  pages   = {633--638},
  year    = {2022},
  doi     = {10.1038/s41567-022-01556-5},
  url     = {https://doi.org/10.1038/s41567-022-01556-5}
}

@article{Ghiotto2021,
  author  = {Ghiotto,  Augusto and Shih,  En-Min and Pereira,  Giancarlo S. S. G. and Rhodes,  Daniel A. and Kim,  Bumho and Zang,  Jiawei and Millis,  Andrew J. and Watanabe,  Kenji and Taniguchi,  Takashi and Hone,  James C. and Wang,  Lei and Dean,  Cory R. and Pasupathy,  Abhay N.},
  title   = {Quantum criticality in twisted transition metal dichalcogenides},
  journal = {Nature},
  volume  = {597},
  pages   = {345--349},
  year    = {2021},
  doi     = {10.1038/s41586-021-03815-6},
  url     = {https://doi.org/10.1038/s41586-021-03815-6}
}

@article{Greene2020,
  author  = {Greene,  Richard L. and Mandal,  Pampa R. and Poniatowski,  Nicholas R. and Sarkar,  Tarapada},
  title   = {The Strange Metal State of the Electron-Doped Cuprates},
  journal = {Annu. Rev. Condens. Matter Phys.},
  volume  = {11},
  pages   = {213--229},
  year    = {2020},
  doi     = {10.1146/annurev-conmatphys-031119-050558},
  url     = {https://doi.org/10.1146/annurev-conmatphys-031119-050558}
}

@article{pnas.1915224117,
  author  = {Connie H. Mousatov  and Erez Berg  and Sean A. Hartnoll},
  title   = {Theory of the strange metal {Sr$_3$Ru$_2$O$_7$}},
  journal = {Proc. Natl. Acad. Sci. U.S.A.},
  volume  = {117},
  pages   = {2852--2857},
  year    = {2020},
  doi     = {10.1073/pnas.1915224117},
  url     = {https://doi.org/10.1073/pnas.1915224117}
}

@article{RevModPhys.79.1015,
  author  = {L\"ohneysen, Hilbert v. and Rosch, Achim and Vojta, Matthias and W\"olfle, Peter},
  title   = {{Fermi}-liquid instabilities at magnetic quantum phase transitions},
  journal = {Rev. Mod. Phys.},
  volume  = {79},
  pages   = {1015--1075},
  year    = {2007},
  doi     = {10.1103/RevModPhys.79.1015},
  url     = {https://doi.org/10.1103/RevModPhys.79.1015}
}

@article{2fcg-zmwt,
  author  = {Thuillier, Davis and Ghosh, Sayak and Ramshaw, B. J. and Scaffidi, Thomas},
  title   = {Multipolar {Fermi} surface deformations in {Sr$_2$RuO$_4$} probed by resistivity and sound attenuation: A window into electron viscosity and the collision operator},
  journal = {Phys. Rev. Lett.},
  volume  = {135},
  pages   = {146302},
  year    = {2025},
  doi     = {10.1103/2fcg-zmwt},
  url     = {https://doi.org/10.1103/2fcg-zmwt}
}

@article{science.aaf2487,
  author  = {Jan Zaanen},
  title   = {Electrons go with the flow in exotic material systems},
  journal = {Science},
  volume  = {351},
  pages   = {1026--1027},
  year    = {2016},
  doi     = {10.1126/science.aaf2487},
  url     = {https://doi.org/10.1126/science.aaf2487}
}

@article{PhysRevLett.113.235901,
  author  = {Tomadin, Andrea and Vignale, Giovanni and Polini, Marco},
  title   = {{Corbino} disk viscometer for {2D} quantum electron liquids},
  journal = {Phys. Rev. Lett.},
  volume  = {113},
  pages   = {235901},
  year    = {2014},
  doi     = {10.1103/PhysRevLett.113.235901},
  url     = {https://doi.org/10.1103/PhysRevLett.113.235901}
}

@article{PhysRevB.95.075127,
  author  = {Eberlein, Andreas and Patel, Aavishkar A. and Sachdev, Subir},
  title   = {Shear viscosity at the {Ising}-nematic quantum critical point in two-dimensional metals},
  journal = {Phys. Rev. B},
  volume  = {95},
  pages   = {075127},
  year    = {2017},
  doi     = {10.1103/PhysRevB.95.075127},
  url     = {https://doi.org/10.1103/PhysRevB.95.075127}
}

@book{landau2013course,
  author    = {Landau, L. D. and Lifshitz, E. M.},
  title     = {Fluid Mechanics},
  series    = {Course of Theoretical Physics},
  volume    = {6},
  edition   = {2},
  publisher = {Pergamon Press},
  address   = {Oxford},
  year      = {1987},
  isbn      = {978-0-7506-2767-2},
  url       = {https://shop.elsevier.com/books/fluid-mechanics/landau/978-0-08-057073-0}
}

@article{science.aad0201,
  author  = {D. A. Bandurin  and I. Torre  and R. Krishna Kumar  and M. Ben Shalom  and A. Tomadin  and A. Principi  and G. H. Auton  and E. Khestanova  and K. S. Novoselov  and I. V. Grigorieva  and L. A. Ponomarenko  and A. K. Geim  and M. Polini},
  title   = {Negative local resistance caused by viscous electron backflow in graphene},
  journal = {Science},
  volume  = {351},
  pages   = {1055--1058},
  year    = {2016},
  doi     = {10.1126/science.aad0201},
  url     = {https://doi.org/10.1126/science.aad0201}
}

@article{science.aad0343,
  author  = {Jesse Crossno  and Jing K. Shi  and Ke Wang  and Xiaomeng Liu  and Achim Harzheim  and Andrew Lucas  and Subir Sachdev  and Philip Kim  and Takashi Taniguchi  and Kenji Watanabe  and Thomas A. Ohki  and Kin Chung Fong},
  title   = {Observation of the {Dirac} fluid and the breakdown of the {Wiedemann--Franz} law in graphene},
  journal = {Science},
  volume  = {351},
  pages   = {1058--1061},
  year    = {2016},
  doi     = {10.1126/science.aad0343},
  url     = {https://doi.org/10.1126/science.aad0343}
}

@article{levitov2016electron,
  author  = {Levitov, Leonid and Falkovich, Gregory},
  title   = {Electron viscosity, current vortices and negative nonlocal resistance in graphene},
  journal = {Nat. Phys.},
  volume  = {12},
  pages   = {672--676},
  year    = {2016},
  doi     = {10.1038/nphys3667},
  url     = {https://doi.org/10.1038/nphys3667}
}

@article{PhysRevLett.86.5986,
  author  = {Lupien, C. and MacFarlane, W. A. and Proust, Cyril and Taillefer, Louis and Mao, Z. Q. and Maeno, Y.},
  title   = {Ultrasound attenuation in {Sr$_2$RuO$_4$}: An angle-resolved study of the superconducting gap function},
  journal = {Phys. Rev. Lett.},
  volume  = {86},
  pages   = {5986--5989},
  year    = {2001},
  doi     = {10.1103/PhysRevLett.86.5986},
  url     = {https://doi.org/10.1103/PhysRevLett.86.5986}
}

@article{PhysRevLett.106.096401,
  author  = {Mravlje, Jernej and Aichhorn, Markus and Miyake, Takashi and Haule, Kristjan and Kotliar, Gabriel and Georges, Antoine},
  title   = {Coherence--incoherence crossover and the mass-renormalization puzzles in {Sr$_2$RuO$_4$}},
  journal = {Phys. Rev. Lett.},
  volume  = {106},
  pages   = {096401},
  year    = {2011},
  doi     = {10.1103/PhysRevLett.106.096401},
  url     = {https://doi.org/10.1103/PhysRevLett.106.096401}
}

@misc{hunter2025,
  author        = {A. Hunter and C. Putzke and F. B. Kugler and S. Beck and E. Cappelli and F. Margot and M. Straub and Y. Alexanian and J. Teyssier and A. de la Torre and K. W. Plumb and M. D. Watson and T. K. Kim and C. Cacho and N. C. Plumb and M. Shi and M. Radovic and J. Osiecki and C. Polley and D. A. Sokolov and A. P. Mackenzie and E. Berg and A. Georges and P. J. W. Moll and A. Tamai and F. Baumberger},
  title         = {Non-{Fermi}-liquid quasiparticles in strain-tuned {Sr$_2$RuO$_4$}},
  year          = {2025},
  eprint        = {2503.11311},
  archivePrefix = {arXiv},
  primaryClass  = {cond-mat.str-el},
  url           = {https://arxiv.org/abs/2503.11311}
}

@article{pnas.2321665121,
  author  = {LingNan Wei  and Qiaoling Xu  and Yangchen He  and Qingxin Li  and Yan Huang  and Wang Zhu  and Kenji Watanabe  and Takashi Taniguchi  and Martin Claassen  and Daniel A. Rhodes  and Dante M. Kennes  and Lede Xian  and Angel Rubio  and Lei Wang},
  title   = {Linear resistivity at van Hove singularities in twisted bilayer {WSe$_2$}},
  journal = {Proc. Natl. Acad. Sci. U.S.A.},
  volume  = {121},
  pages   = {e2321665121},
  year    = {2024},
  doi     = {10.1073/pnas.2321665121},
  url     = {https://doi.org/10.1073/pnas.2321665121}
}

@article{PhysRevB.100.035135,
  author  = {Ye, Mengxing and Chubukov, Andrey V.},
  title   = {{Hubbard} model on a triangular lattice: Pseudogap due to spin-density-wave fluctuations},
  journal = {Phys. Rev. B},
  volume  = {100},
  pages   = {035135},
  year    = {2019},
  doi     = {10.1103/PhysRevB.100.035135},
  url     = {https://doi.org/10.1103/PhysRevB.100.035135}
}

@article{PhysRevLett.110.127001,
  author  = {Wang, Yuxuan and Chubukov, Andrey V.},
  title   = {Superconductivity at the Onset of Spin-Density-Wave Order in a Metal},
  journal = {Phys. Rev. Lett.},
  volume  = {110},
  pages   = {127001},
  year    = {2013},
  doi     = {10.1103/PhysRevLett.110.127001},
  url     = {https://doi.org/10.1103/PhysRevLett.110.127001}
}

@article{PhysRevLett.130.126001,
  author  = {Wu, Yi-Ming and Wu, Zhengzhi and Yao, Hong},
  title   = {Pair-Density-Wave and Chiral Superconductivity in Twisted Bilayer Transition Metal Dichalcogenides},
  journal = {Phys. Rev. Lett.},
  volume  = {130},
  pages   = {126001},
  year    = {2023},
  doi     = {10.1103/PhysRevLett.130.126001},
  url     = {https://doi.org/10.1103/PhysRevLett.130.126001}
}

@article{PhysRevLett.83.3538,
  author  = {Varma, C. M.},
  title   = {Pseudogap Phase and the Quantum-Critical Point in Copper-Oxide Metals},
  journal = {Phys. Rev. Lett.},
  volume  = {83},
  pages   = {3538--3541},
  year    = {1999},
  doi     = {10.1103/PhysRevLett.83.3538},
  url     = {https://doi.org/10.1103/PhysRevLett.83.3538}
}

@article{PhysRevLett.114.147001,
  author  = {Benhabib, S. and Sacuto, A. and Civelli, M. and Paul, I. and Cazayous, M. and Gallais, Y. and M\'easson, M.-A. and Zhong, R. D. and Schneeloch, J. and Gu, G. D. and Colson, D. and Forget, A.},
  title   = {Collapse of the normal-state pseudogap at a {Lifshitz} transition in the {Bi$_2$Sr$_2$CaCu$_2$O$_{8+\delta}$} cuprate superconductor},
  journal = {Phys. Rev. Lett.},
  volume  = {114},
  pages   = {147001},
  year    = {2015},
  doi     = {10.1103/PhysRevLett.114.147001},
  url     = {https://doi.org/10.1103/PhysRevLett.114.147001}
}

@article{doiron2017pseudogap,
  author  = {Doiron-Leyraud, N and Cyr-Choini{\`e}re, O and Badoux, S and Ataei, A and Collignon, C and Gourgout, A and Dufour-Beaus{\'e}jour, S and Tafti, FF and Lalibert{\'e}, F and Boulanger, M-E and others},
  title   = {Pseudogap phase of cuprate superconductors confined by {Fermi} surface topology},
  journal = {Nat. Commun.},
  volume  = {8},
  pages   = {2044},
  year    = {2017},
  doi     = {10.1038/s41467-017-02122-x},
  url     = {https://doi.org/10.1038/s41467-017-02122-x}
}

@article{yuan2022scaling,
  author  = {Yuan, Jie and Chen, Qihong and Jiang, Kun and Feng, Zhongpei and Lin, Zefeng and Yu, Heshan and He, Ge and Zhang, Jinsong and Jiang, Xingyu and Zhang, Xu and others},
  title   = {Scaling of the strange-metal scattering in unconventional superconductors},
  journal = {Nature},
  volume  = {602},
  pages   = {431--436},
  year    = {2022},
  doi     = {10.1038/s41586-021-04305-5},
  url     = {https://doi.org/10.1038/s41586-021-04305-5}
}

@article{jiang2023interplay,
  author  = {Jiang, Xingyu and Qin, Mingyang and Wei, Xinjian and Xu, Li and Ke, Jiezun and Zhu, Haipeng and Zhang, Ruozhou and Zhao, Zhanyi and Liang, Qimei and Wei, Zhongxu and others},
  title   = {Interplay between superconductivity and the strange-metal state in {FeSe}},
  journal = {Nat. Phys.},
  volume  = {19},
  pages   = {365--371},
  year    = {2023},
  doi     = {10.1038/s41567-022-01894-4},
  url     = {https://doi.org/10.1038/s41567-022-01894-4}
}

@article{PhysRevE.108.044101,
  author  = {Zaccone, Alessio},
  title   = {General theory of the viscosity of liquids and solids from nonaffine particle motions},
  journal = {Phys. Rev. E},
  volume  = {108},
  pages   = {044101},
  year    = {2023},
  doi     = {10.1103/PhysRevE.108.044101},
  url     = {https://doi.org/10.1103/PhysRevE.108.044101}
}
\end{document}